\documentclass[12pt,a4paper]{article}
\usepackage{amssymb}
\usepackage{amsmath, amsthm,amssymb,epsfig,latexsym}
\usepackage{amsopn}
\usepackage{pifont}
\usepackage{color}
\usepackage{easybmat}
\usepackage{mathrsfs}
\usepackage{ulem}
\usepackage[numbers,square,sort&compress]{natbib}
\newcommand{\so}{\scriptscriptstyle \rm I}
\newcommand{\st}{\scriptscriptstyle \rm I\hspace{-1pt}I}
\newcommand{\sth}{\scriptscriptstyle \rm I\hspace{-1pt}I\hspace{-1pt}I}
\newcommand{\down}{\downarrow}
\newcommand{\up}{\uparrow}
\newcommand{\uc}{u^{\scriptscriptstyle C}}
\newcommand{\ub}{u^{\scriptscriptstyle B}}
\newcommand{\vc}{v^{\scriptscriptstyle C}}
\newcommand{\vb}{v^{\scriptscriptstyle B}}

\newcommand{\bu}{\bar u}
\newcommand{\bv}{\bar v}
\newcommand{\bx}{\bar x}
\newcommand{\by}{\bar y}

\newcommand{\bY}{\bar Y}
\newcommand{\bZ}{\bar Z}

\newcommand{\buc}{\bar{u}^{\scriptscriptstyle C}}
\newcommand{\bub}{\bar{u}^{\scriptscriptstyle B}}
\newcommand{\bvc}{\bar{v}^{\scriptscriptstyle C}}
\newcommand{\bvb}{\bar{v}^{\scriptscriptstyle B}}

\newcommand{\Qone}{Q^{(1)}}
\newcommand{\Qtwo}{Q^{(2)}}

\newcommand{\bw}{\bar w}

\newcommand{\be}[1]{\begin{equation}\label{#1}}
\newcommand{\ba}[1]{\begin{multline}\label{#1}}
\newcommand{\ee}{\end{equation}}
\newcommand{\ea}{\end{eqnarray}}

\newcommand{\diag}{\mathop{\rm diag}}

\newcommand{\str}{\mathop{\rm str}}

\newcommand{\Fin}{{\mathop{\rm Fin}}}
\newcommand{\Finite}{{\mathop{\rm Finite\;\; part}}}
\newtheorem{thm}{Theorem}[section]
\newtheorem{prop}{Proposition}[section]

\def\qed{\hfill\nobreak\hbox{$\square$}\par\medbreak}
 \makeatletter
 \@addtoreset{equation}{section}
 \makeatother
 
\newcommand{\bea}{\begin{eqnarray}}
\newcommand{\eea}{\end{eqnarray}}

\newcommand{\ket}[1]{{\left|#1\right\rangle}}
\newcommand{\bra}[1]{{\left\langle #1\right|}}
\newcommand{\vev}[1]{\left\langle #1 \right\rangle}

\usepackage{ifpdf}

\ifpdf
\usepackage{epstopdf}
\usepackage[pdftex,colorlinks,urlcolor=blue,citecolor=blue,linkcolor=blue]{hyperref}
\else
\usepackage[hypertex,colorlinks,urlcolor=blue,citecolor=blue,linkcolor=blue]{hyperref}
\fi
\begin{document}

\thispagestyle{empty}

\begin{flushright}
\hphantom{111111111111}
\end{flushright}

\vspace{12pt}

\begin{center}
\begin{LARGE}
  {\bf  The LeClair-Mussardo series }

  \medskip
  
  {\bf and nested Bethe Ansatz}
\end{LARGE}

\vspace{40pt}

{A.~Hutsalyuk${}^{a}$, B.~Pozsgay${}^{b,c}$,
L.~Pristy\'ak${}^{d}$\  \footnote{
hutsalyuk@gmail.com, pozsgay.balazs@gmail.com, levente.pristyak@gmail.com}}

\vspace{10mm}

${}^a$ {\it BME ``Momentum" Statistical Field Theory Research Group,\\
Department of Theoretical Physics,\\
Budapest University of Technology and Economics %,
% 1521 Budapest, Hungary
}

${}^b$ {\it
  Department of Theoretical Physics, \\
  E\"otv\"os Lor\'and University Budapest
}

${}^c$ {\it
  MTA-ELTE ``Momentum'' Integrable Quantum Dynamics Research Group,\\
  E\"otv\"os Lor\'and University Budapest
}

${}^d$ {\it Department of Theoretical Physics, \\
  Budapest University of Technology and Economics\\
  % 1521 Budapest, Hungary
}

\vspace{2mm}

\end{center}

\vspace{4mm}

\begin{abstract}
We consider correlation functions in one dimensional quantum
  integrable models related to the algebra symmetries
  $\mathfrak{gl}(2|1)$ and $\mathfrak{gl}(3)$.  Using the algebraic
  Bethe Ansatz approach we develop {an expansion theorem, which
    leads to an infinite
  integral series  in the thermodynamic limit}. The series is the generalization of the
  LeClair-Mussardo series {to nested Bethe Ansatz systems}, 
    and it is applicable both to one-point and two-point
    functions.
     As an example we consider the
  ground state density-density correlator in the Gaudin-Yang model of
  spin-1/2 Fermi particles. Explicit formulas are presented in a
  special large coupling and large imbalance limit.
\end{abstract}

\begin{center}

\today

\end{center}

\thispagestyle{empty}

\setcounter{footnote}{0}

\section{Introduction}

Correlation functions in the theory of integrable models have been an object of interest for many
years. The study of them goes back to the pioneering works \cite{Izergin1984,Izergin1985,Korepin1,CTW1,CTW2,Honerkamp,KorepinSlavnov,Slavnov90}.
For non-relativistic systems the established canonical framework is the 
Quantum Inverse Scattering Method (QISM) \cite{Faddeev2,KULISH1981246,KulRes82,kulish-resh-glN}, which led to a study
of correlation functions in various models such as the Heisenberg spin chains and the 1D Bose gas,
see for example \cite{KitanineMaiilletTerrasSlavnov2002,maillet-dynamical-1,KitanineMaiilletTerrasSlavnov2008,formfactor-asympt,KitMT00,GohKS04}. In these models finite temperature \cite{Gohmann1,Dugave_2014, KozlowskiMailletSlavnov1,karol-corr2,Gohmann2,dugave2014formfactor,frank-karol-junji-maxim-asympt-GS-2} and dynamical correlation functions
\cite{KitanineMailletSlavnovTerras2005,KozlowskiMailletSlavnov4,KozlowskiTerras,Gohmann3} were also derived in recent
years. The main strategy of these works is to embed the local operators of the system into the
Yang-Baxter algebra, and to derive the form factors and correlation functions using Algebraic Bethe
Ansatz (ABA) \cite{BIK}.

Different methods have been developed for integrable relativistic Quantum Field Theories. 
In these models the form factors
are obtained as solutions to the so-called form factor bootstrap
program
\cite{Karowski:1978vz,Smirnov-Book,zam_Lee_Yang,Delfino:1996nf}.
Originally the form factor approach was developed to compute
two-point functions at zero temperature, but later it was found that even finite temperature
correlations can be obtained with it. A key development was an integral series derived for finite
temperature one-point functions in \cite{LM}, which is today known as the
LeClair-Mussardo (LM) series. The validity of the LM series for two-point functions was questioned first
in \cite{Saleur,CastroAlvaredo}, and it was pointed out in \cite{D22} that it is not
even well defined. Later a well defined prescription for the two-point functions was found in
\cite{Pozsgay3}. Interestingly, in the case of the 1D Bose gas the LM series was discovered already in the very early
works \cite{CTW1,CTW2}, which treated static two-point functions. These early results were partially
forgotten and not discussed in most of the papers dealing with the LM
series. 

Most of the theoretical work (both on the spin chains and in field
theory) focused on ground state and finite
temperature correlations. However, the last 10 years saw tremendous progress in the field of 
non-equilibrium dynamics and quantum quenches in particular, and this led to the study of
Generalized Gibbs Ensembles (GGE's), which arise as steady states in
these systems \cite{rigol-gge,essler-fagotti-quench-review}. It is now generally accepted that in Bethe
Ansatz solvable systems the information stored in a GGE is equivalent to specifying the rapidity
distributions of the quasi-particles \cite{JS-CGGE}, and this motivated the study of correlation
functions in Bethe states with arbitrary root distributions \cite{Pozsgay2,sajat-corr}. Such studies
were further spurred by the development of Generalized Hydrodynamics (GHD)
\cite{doyon-GHD,jacopo-GHD}, which describes the large scale transport properties of integrable
models.  A central assumption in GHD is the existence of fluid cells such that each cell is
described by a space and time dependent GGE, thus the rapidity distributions can become complicated
dynamical functions. An especially interesting result for dynamical two point functions
was derived in \cite{Doyon-GHD-LM}, using the GHD and the LM series.

Despite all of the progress described above, most of these works are concerned with models of one-component gases or
spin chains related to $\mathfrak{gl}(2)$ symmetry. Much less is known about the correlation functions of
spin chains with $\mathfrak{gl}(N)$ spin symmetry and multi-component gases. There are partial
results for short-range correlation functions in higher rank spin chains
\cite{kluemper-su3,boos-artur-qkz-su3,ribeiro2020correlation}. {Furthermore, form factors of
  local operators 
  were computed in models related to the algebra symmetry $\mathfrak{gl}(3)$ and
  $\mathfrak{gl}(2|1)$ in \cite{Pozsgay2012,Wheeler,PRS2012,PRS2015}
  using the nested ABA approach. This allowed to treat the asymptotic
  behavior of correlation function using form factor expansion
  \cite{Karol-Ragoucy-nested-asympt}.} On the other hand, up to now
there have been no results
for generic correlation functions at small or intermediate distances,
and it was not known whether some
generalization of the LM series exists in the nested models.

Our goal is to contribute to the solution of these problems. We develop a form factor expansion in
nested Bethe Ansatz systems, which can be applied to one-point and (static) two-point functions in
arbitrary equilibrium ensembles. Thus it can be used both for the
ground state, in Gibbs and Generalized Gibbs Ensembles, but also in non-equilibrium
situations where the GHD can be applied.
Our result is the generalization of the LeClair-Mussardo formula to
nested non-relativistic models. We use standard
tools and also recent developments of the algebraic Bethe Ansatz to
derive the LM series. We also explain how to compute the so-called
symmetric form factors of the two-point functions, which are the key
ingredients of the LM series.  

The method is applicable for an arbitrary model related to algebra symmetry $\mathfrak{gl}(2|1)$
or $\mathfrak{gl}(3)$.
We focus on one specific example, namely the 1D spin-1/2 Fermi gas
(a.k.a. Gaudin-Yang model, which is related to the algebra symmetry $\mathfrak{gl}(2|1)$). A similar
example of the two-component 1D 
Bose gas will be considered separately.

The paper is organized as follows. In Section \ref{results} the main results of the paper are described,
omitting the technical details in the various definitions. Notation
and short description of Bethe Ansatz are given. In section 
\ref{LMseries} we introduce the symmetric form factors and prove {
the finite volume expansion theorem} for algebra symmetry
$\mathfrak{gl}(2|1)$ related models. The case of algebra symmetry
$\mathfrak{gl}(3)$ related models can be obtain in a similar
way and the difference is minor between these cases. In Section
\ref{thermodynamic_limit} the thermodynamic limit of 
the {form factor expansion} is considered. In Section  \ref{correlation_functions} we present an efficient way
 for the computation of the symmetric form factors, focusing on the
 Gaudin-Yang model. Finally, this Section includes our explicit
 results for the correlators in the combined large coupling and large
 imbalance limits.
Appendix \ref{quasiReshet} includes technical details of the computations involving  the symmetric
form factors. In Appendix \ref{Charge_current} we consider the mean
values and form factors of the conserved charges, and we perform a
check of the expansion theorem. Finally, in Appendix  \ref{LM_vs_IK}
the comparison of the the integral series with result of
\cite{Korepin1} for the correlation functions is given, considering
the example of 1D Bose gas (Lieb-Liniger). 
 
\section{Main results\label{results}}

In this Section we summaries our main results and apply them to the one dimensional two component quantum gas
models. Most of our results for correlation functions are valid in a much more general setting: they can be applied in
Yang-Baxter integrable models related to the symmetry algebras $\mathfrak{gl}(3)$ and
$\mathfrak{gl}(2|1)$. This will be explained in later Sections. However, in this summary we focus on
the concrete applications. 

Therefore we consider the spin-1/2 one-dimensional $\delta$-function interacting gas models (Gaudin-Yang model) defined through the Hamiltonian \cite{Gaudin, Yang} (see also \cite{YangSutherland, Sutherland75, McGuire}).
\be{GY_Hamiltonian}
H=\int_0^Ldx \left\{\partial\psi_{\alpha}^{\dagger}\partial\psi_{\alpha}+2c\psi_{\alpha}^{\dagger}\psi_{\beta}^{\dagger}\psi_{\beta}\psi_{\alpha}\right\},\qquad {\alpha, \beta =\uparrow,\downarrow},
\ee
with canonical Fermi fields satisfying the exchange relations
$\left\{\psi_{\alpha}^{\dagger}(x),\psi_{\beta}(y)\right\}=\delta_{\alpha\beta}\delta(x-y)$.  As
explained below, this model is related to the algebra $\mathfrak{gl}(2|1)$. In \eqref{GY_Hamiltonian} automatic summation over
$\alpha,\beta=\uparrow,\downarrow$ 
is understood. 

Similarly, we can also consider the two component Bose gas defined through a Hamiltonian which has a
form identical to \eqref{GY_Hamiltonian}  but with bosonic fields
$\psi_\uparrow(x),\psi_\downarrow(x)$ satisfying commutation relations 
$\left[\psi_{\alpha}^{\dagger}(x),\psi_{\beta}(y)\right]=\delta_{\alpha\beta}\delta(x-y)$.
The bosonic model is related to the algebra $\mathfrak{gl}(3)$. In both cases periodic boundary conditions are assumed.

Both models can be solved by the nested Bethe Ansatz \cite{Yang, YangSutherland,
  TakahashiGas,KULISH1981246,KulRes82}. Eigenstates  can be characterized by two sets of rapidities
$\bv=\{v_1,\dots,v_b\}$ and $\bu=\{u_1,\dots,u_a\}$.
The  first set of the rapidities describes the momentum of the particles, and the second set describes
the orientation of the wave function in the internal spin space.

Let us however note that in further Sections we call the spin
  variables $\bu$ the first level and rapidity variables $\bv$ the second level. The only consequence of
  our choice will be the unusual convention for the numeration of levels. To summarize our conventions: the first set $\bu$ describes the spin degrees of freedom, and the second set $\bv$ describes the physical momenta of the particles.

We will focus mostly on the fermionic case. In this model the rapidities satisfy the Bethe equations
\cite{TakahashiGas}: 
\be{betheGY}
\begin{split}
&e^{i p(v_j)L}=\prod_{s=1}^a S_{12}(v_j,u_s),\qquad j=1,\dots,b,\\
&1=\prod_{s=1}^aS(u_j,u_s)\prod_{\ell=1}^bS_{12}(v_{\ell},u_j),\qquad j=1,\dots,a,
\end{split}
\ee
where $S$, $S_{12}$ are phases of (quasi)particles scattering 
\be{scattering}
S_{12}(x)=\frac{x+ic/2}{x-ic/2},\qquad S(x)=\frac{x+ic}{x-ic},
\ee
and  $p(v)$ is the single particle momentum, which in this model is simply $p(v)=v$. For the
  Bethe equations in the bosonic case see \cite{KULISH1981246, Yang}. 

In this work we are interested in static two point correlation functions in these models. We focus on 
the density-density correlation functions
\be{densities_correlator}
\vev{O(x,y)}=\vev{q_\alpha(x) q_\beta(y)},\qquad  q_\alpha(x)=\psi_\alpha^\dagger(x)\psi_\alpha(x),\qquad \alpha=\uparrow,\downarrow
\ee
and similarly in the bosonic case. {Furthermore, in the actual
computations we will consider the correlators
\begin{equation}
  \vev{q_\up(x)q_\down(y)},\qquad \vev{q(x)q(y)},
\end{equation}
where $q(x)=q_\up(x)+q_\down(x)$ is the total density operator.}

In the following we present expansion theorems for the mean
values in both finite and infinite volumes; in order to simplify the notations we will denote the
operator product as $O(x,y)$. Specific details of the local operators will enter the form factors involved.

Our first main result is the following finite volume expansion theorem:
\be{LM}
\langle O(x,y)\rangle_{a,b}\left(\bu;\bv\right)=
\frac{\sum \mathfrak F_{\; a-m,b-n}\left(\bu_{\st};\bv_{\st}\right)
\det G\left(\bu_{\so},\bv_{\so}\right)}
{\det G\left(\bu,\bv\right)}.
\ee

Here $\langle O(x,y)\rangle$ denotes the normalized expectation value
and the sum on the r.h.s. is taken over all bipartite partitions $\bu\to\{\bu_{\so},\bu_{\st}\}$,
$\bv\to\{\bv_{\so},\bv_{\st}\}$\footnote{We should stress that here and further we use partitions where the cases $\#\bu_{\so}=0$, $\#\bu_{\st}=\#\bu=a$ or vice versa (and the same for set $\bv$) are included.}, $\#\bu=a$, $\#\bv$=b. The matrix $G$ is the so-called {\it Gaudin matrix}, which is the
Jacobian of the Bethe equations of the system with some finite size
$L$. The cardinalities of the sets 
are $\#\bu_{\so}=m$, $\#\bv_{\so}=n$ and $\#\bu_{\st}=a-m$, $\#\bv_{\st}=b-n$. The object $\mathfrak F_{m,n}$
is the so-called {\it symmetric form factor} of the composite operator, which is a special limit of
an off-diagonal form factor defined as
\be{symm_limit}
\mathfrak F_{m,n}(\bu;\bv)=\mathcal N^{-1}\lim_{\varepsilon\to 0}\langle \bu+\varepsilon;\bv+\varepsilon|O(x,y)|\bu;\bv\rangle.
\ee
Precise details of this definition, including the normalization $\mathcal N$ are presented in
Section \ref{expansion}. We should note that in a particular model the symmetric form factors
$\mathfrak F_{a,b}$ can be non-zero even if there is no Bethe state
corresponding to the particle numbers $a,b$. This happens because the
symmetric form factor describes the amplitude of a multi-particle process {\it in the
  presence} of the other particles, and thus the selection rules for
them can be different from those for the states.

This type of finite volume form factors expansion was first used in  \cite{Takacs1} for generic local operators in
integrable QFT. It was later shown in \cite{Pozsgay2} that in the thermodynamic limit this expansion
reproduces the so-called LeClair-Mussardo series \cite{LM}, which is an integral series for the mean
values. It was argued in \cite{Pozsgay3} that in QFT the expansion theorem should hold also for composite
objects such as the two-point functions. We should also note that in spin chains  related to algebra
symmetry $\mathfrak{gl}(2)$ the theorem was proven rather generally in \cite{yunfeng1,Pozsgay1}.

The expansion theorem treats the operator product as a single object, and the form factors depend on
the operators and the distance $x-y$. In this respect our method is different from the direct
approaches, where a complete set of states is inserted between the two operators.
The advantage of
this method is that in the thermodynamic limit it leads to integral series involving form factors
with low particle numbers, which can be determined from relatively simple computations.

The resulting integral series can be considered the LeClair-Mussardo series for nested Bethe Ansatz
models, which we describe in the following.
Let us consider for simplicity a situation where { in ground state} the solutions to the Bethe equations are purely
real. Extensions to the general case with so-called string solutions will be considered
elsewhere. We take the thermodynamic limit of the formula \eqref{LM} such that {the particle
densities are kept finite. Then we obtain the following result:
\be{LeClairMussardoPozsgayHutsaliuk}
\langle O(x,y)\rangle=\sum_{m,n=0}^{\infty}\frac{1}{m!n!}\prod_{j=1}^m \int\frac{dt_j}{(2\pi)} \omega^{(1)}(t_j)
\prod_{i=1}^n
\int \frac{ds_i}{(2\pi)}\;\omega^{(2)}(s_i) \mathfrak F_{m,n}\left(\bar t;\bar s\right),
\ee
where $\omega^{(1,2)}$ are weight functions defined in \eqref{thermo_ratio_Gaudin}.
This series can be considered {as the generalization of the LeClair-Mussardo formula, and it is proven
in Section \ref{expansion_thermo}}. 

The series involves the
``symmetric'' diagonal form factors. In contrast, the original LeClair-Mussardo formula used a
different prescription for the diagonal limit, which are called the
``connected'' diagonal form factors. Furthermore, the original LM series does not
involve the additional $\omega$ weight functions. The
equivalence of the two integral series follows from relations between the two different evaluations of the
diagonal limit, as discussed in length in \cite{Takacs1}.
We should note that in the one-component
case the LM series for two-point functions was already derived in \cite{CTW1,CTW2}, and these works
also used the connected form factors.

We also present the LM series in the nested case using the connected form factors, which is written
as
\be{LeClairMussardoPozsgayHutsaliukConnected}
\langle O(x,y)\rangle=\sum_{m,n=0}^{\infty}\frac{1}{m!n!}\prod_{j=1}^m \int\frac{dt_j}{(2\pi)} 
\prod_{i=1}^n
\int \frac{ds_i}{(2\pi)}\; \mathfrak F^c_{m,n}\left(\bar t;\bar s\right).
\ee
Here $\mathfrak F^c_{m,n}\left(\bar t;\bar s\right)$ are the connected form factors, defined as
\be{connected_lim}
\mathfrak F^c_{a,b}(\bu;\bv)=\Finite\left\{\mathcal N^{-1}\lim_{\bar\varepsilon\to0,\;\bar\varepsilon'\to0}\langle \bu+\bar\varepsilon;\bv+\bar\varepsilon'|O(x,y)|\bu;\bv\rangle\right\}.
\ee
Precise details of this definition are presented in Section \ref{sec:connected}.

Using \eqref{LM} with \eqref{symm_limit} the static zero-temperature two-point correlation functions can be calculated.
We apply them to a specific large imbalance limit of the two-component fermionic
model, which we describe in the following.

In the two-component Fermi gas with $c>0$ there are two natural small parameters, which can be
chosen as $Q/c$ and $B/c$, where $Q$ and $B$ are the Fermi rapidities in the ground state, which
involves only real roots. These two
parameters can be related to the particle densities, the exact formulas are presented later in Section
\ref{thermodynamic_limit}. 
Smallness of $Q/c$  means large coupling limit, whereas
smallness of $B/c$ is equivalent to large imbalance, or smallness off the ratio 
$n_\down/n$, where $n=\langle q(x)\rangle$, $n_{\uparrow,\downarrow}=\langle
  q_{\uparrow,\downarrow}(x)\rangle$. If the parameter $B/c$ is small, then the densities can be expressed
as
\be{BQ_vs_densities}
\begin{split}
n_{\down}&=\vev{q_\down(x)}=2B  \frac{1}{(2\pi)^2}  4 \arctan(2Q/c)+O(B^2/c^2),\\
n&=\vev{q(x)}=\frac{Q}{\pi}+\frac{4B}{\pi^3} \arctan^2(2Q/c)+O(B^2/c^2).
\end{split}
\ee
Note that these formulas are exact in $Q/c$. However in the case of two-point correlation functions the series \eqref{LeClairMussardoPozsgayHutsaliuk}  converges only when the ratio $Q/c$ is small too. Thus we will develop a Taylor series  in the two small parameters.

Our result for the leading terms of the correlation functions is 
\be{corr_density}
\begin{split}
\langle
q(x)q(0)\rangle=\left(\frac{Q^2}{\pi^2}-\frac{\sin^2(Qx)}{x^2\pi^2}\right)\times
\left(1+O(B^2/c^2)+O(Q^2/c^2)+O(QB/c^2)\right).
\end{split}
\ee
\be{corr_mixed}
\begin{split}
\langle
q_{\uparrow}(x)q_{\downarrow}(0)\rangle=\left(\frac{4Q^2B}{\pi^3c}-\frac{4B\sin^2(Qx)}{\pi^3x^2c}\right)\times
\left(1+O(B/c)+O(Q/c)\right).
\end{split}
\ee

These leading terms describe free fermionic behavior in  both cases. We expect that this behavior only holds at the particular level
of approximation. Note that the correction terms to the formula \eqref{corr_density} are quadratic; the absence of linear terms (carrying an additional factor of $B/c$ or $Q/c$) is a non-trivial result of our computations.}

In the $x\to 0$ limit both formulas above approach zero. This agrees with result of Section \ref{densities} where  leading terms of the local correlators were found from the Hellmann-Feynman theorem. It is shown there that
\be{Hellmann_estimation}
\vev{q(0)q(0)}= 2\vev{q_\up(0)q_\down(0)}\sim \frac{Q^4B}{c^3}+O(c^{-4}).
\ee
Clearly, this is sub-leading compared to the result \eqref{corr_mixed}.

\subsection{Notations\label{Notations}}

In order to simplify the formulas we use the following notational conventions in the paper.

We denote  sets of arbitrary variables as $\bu=\{u_1,u_2,\dots\}$, $\bv=\{v_1,v_2,\dots\}$. {Subsets are labeled by roman or Arabic numbers, e.g. $\bv_{\so}$, $\bv_{\st}$, $\bu_1$, $\bu_2$, etc.} We use  the notation $\bu_j=\bu\setminus u_j$, $\bu_{j,k}=\bu\setminus \{u_j,u_k\}$ for sets with the certain elements are omitting. For an arbitrary function $F(x)$, $F(x,y)$ and  for an arbitrary sets $\bx$, $\by$ we use the notation
\be{single_prod}
\begin{split}
 F(\bar x)=\prod_{j=1}^{\#\bar x} F(x_j),\qquad
F(s,\bar y)=\prod_{j=1}^{\#\bar y}F(s,y_j),\qquad F(\bar x,\bar y)=\prod_{j=1}^{\#\bar
  y}\prod_{k=1}^{\#\bar x}F(x_k,y_j).
\end{split}
\end{equation}
{ In a similar way
\be{complement}
F(\bar y_k)=\prod_{j=1;j\ne k}^{\#\bar y}F(y_j),\qquad F(\bar x_k, \bar y_s)=\prod_{j=1;j\ne k}^{\#\bar x}\prod_{\ell=1;\ell\ne s}^{\#\bar y} F(x_j,y_{\ell}).
\ee
}
Skew-symmetric products of an arbitrary function ${\mathfrak g}(x,y)$ over the set $\bar x$
are defined as
\be{skew_prod}
\Delta_{\mathfrak g}(\bar x)=\prod_{i>j}\mathfrak g(x_i,x_j),\qquad \Delta'_{\mathfrak g}(\bar x)=\prod_{i<j}\mathfrak g(x_i,x_j).
\ee

\subsection{Bethe Ansatz\label{BEandGR}}

We consider models with algebra symmetry $\mathfrak{gl}(3)$ and $\mathfrak{gl}(2|1)$
simultaneously.
We present detailed computations mostly in the case of  $\mathfrak{gl}(2|1)$;
the case of algebra symmetry $\mathfrak{gl}(3)$ is very similar
and we will just give the final results. 

Throughout the paper we use conventional notations for the following functions:
\be{def}
\begin{split}
&f(x,y)=\frac{x-y+ic}{x-y},\qquad
g(x,y)=\frac{ic}{x-y},\qquad 
h(x,y)=\frac{x-y+ic}{ic},\qquad\\
&t(x,y)=\frac{g(x,y)}{h(x,y)}=-\frac{c^2}{(x-y)(x-y+ic)}.
\end{split}
\ee
We will also use the kernel $K(x,y)$ defined as
\be{K}
K(x,y)=\frac{2c^2}{(x-y)^2+c^2}.
\ee

We define a $Z_2$-graded vector space $\mathbb C^{2|1}$ with a grading $[1]=[2]=0$, $[3]=1$\footnote{Here and further square brackets denote parity.}. Matrices acting in $\mathbb C^{2|1}$ in case of $\mathfrak{gl}(2|1)$
algebra symmetry are also graded with grading defined as $[e_{ij}]=[i]+[j]$ where elementary units $e_{ij}$
are defined as $(e_{ij})_{ab}=\delta_{ai}\delta_{bj}$. 

Our computations will be performed using the Algebraic Bethe Ansatz  (ABA) technique. In ABA the
system is given by certain {\it monodromy matrix} $T$ that satisfies the so-called {\it RTT relation}
  \cite{Faddeev2, KulS80, KulRes82, KulishSklyanin} with an appropriate $R$-matrix
\be{RTT}
R_{12}(u-v) T_{01}(u)T_{02}(v)=T_{02}(v)T_{01}(u)R_{12}(u-v),
\ee
that holds in a tensor product of spaces $V_1\otimes V_2\otimes \mathcal H$ (subscripts of $R_{12}$ and $T_{0k}$ denote spaces
in which correspondent matrix acts non-trivially). Here the so-called {\it auxiliary
  spaces} $V_1$ and $V_2$ are $\mathbb C^3$ in case of algebra symmetry $\mathfrak{gl}(3)$ and $\mathbb
C^{2|1}$ in case of algebra symmetry $\mathfrak{gl}(2|1)$ related models. So-called {\it quantum space} $\mathcal H$ 
coincides with the Hilbert space of the Hamiltonian of considering system.

{The cases of rational $\mathfrak{gl}(3)$ and $\mathfrak{gl}(2|1)$-invariant $R$-matrices are considered in this paper}
\be{Rmat}
R(u-v)=\mathbb I+\frac{ic\mathbb P}{u-v},\qquad \mathbb P=\sum_{i,j=1}^3(-1)^{[j]}e_{ij}\otimes e_{ji},\qquad \mathbb I=\sum_{i,j=1}^{3}e_{ii}\otimes e_{jj},
\ee
with a (graded) permutation operator $\mathbb P$ and a unity operator $\mathbb I$ \cite{KulS80}.

{The Gaudin-Yang model, discussed in introduction, is related to algebra symmetry $\mathfrak{gl}(2|1)$ while the two-component Bose gas to algebra symmetry $\mathfrak{gl}(3)$.} In this paper we do not give the explicit form of monodromy matrix entries for the models under consideration, instead we refer to \cite{KULISH1981246,KulRes82}. 

We denote Bethe Ansatz vacuum as $|0\rangle$ (and dual vacuum as $\langle 0|$). Vacuum eigenvalues of the diagonal entries of the monodromy matrix $T_{ii}$\footnote{Here and further subscripts of $T_{ij}$ denote the matrix element indices, not  numbers of spaces.} are denoted by $\lambda_i$:
\be{lambda}
T_{ii}(t)|0\rangle=\lambda_i(t)|0\rangle,\qquad i=1,\dots,3,
\ee
\be{rr}
r_1(t)=\lambda_1(t)/\lambda_2(t),\qquad r_3(t)=\lambda_3(t)/\lambda_2(t).
\ee

  Following \cite{Izergin1984} we use the concept of the two-site (also called partial) model for
  the description of subsystems. Thus we
consider the system of length $L$ and split it into two subsystems $[0,x]$ and $[x,L]$; this will be
useful for the computation of the two-point correlation functions.
We denote the partial quantities belonging to the first (second) subsystem by superscript $(1)$
(correspondingly $(2)$). For example $Q_i^{(1)}$
denotes the number of particles of type $i$, $i=1,2$, in the first subsystem. Furthermore, $r_i^{(k)}(t)$, $i=1,3$,
$k=1,2$ denote the ratios of eigenvalues of the diagonal elements of partial monodromy matrix belonging  to
subsystem $k$.  Obviously $r_i=r_i^{(1)}r_i^{(2)}$, $i=1,3$.
We apply the following notations for logarithmic  derivatives of $r_i(w)$:
\be{zy}
\begin{split}
Z_k=i\partial_{u_k}\log r_1(u_k),\qquad Y_k=i\partial_{v_k}\log r_3(v_k).
\end{split}
\ee

We extend notations $F(\bu)=F(u_1)F(u_2)\dots$ introduced for
functions to operators, that commute among themselves. For instance {in case of the algebra symmetry $\mathfrak{gl}(2|1)$ related models} $T_{12}(u)T_{12}(v)=T_{12}(v)T_{12}(u)$, thus
\be{OP_prod}
T_{12}(\bu)=T_{12}(u_1),\dots T_{12}(u_a), \qquad \#\bu=a.
\ee
Operators $T_{13}$,  $T_{31}$, $T_{32}$ and $T_{23}$ do not commute among themselves. For example
\be{T13T13}
T_{13}(v)T_{13}(u)h(v,u)=T_{13}(u)T_{13}(v)h(u,v).
\ee
In these cases we introduce the operator products
\be{boldT}
\mathbb T_{j3}(\bv)=\Delta_h^{-1}(\bv)T_{j3}(\bv),\qquad \mathbb T_{3j}(\bv)=\Delta_h^{'-1}(\bv)T_{3j}(\bv),\qquad j=1,2,
\ee
that are symmetric with respect to the parameters $\bv$.

{The transfer matrix, defined as
\be{transfer}
t(u)=\str T=\sum_{i=1}^3 (-1)^{[i]}T_{ii}(u),
\ee
is a generating function of (mutually commuting) integrals of motion $\{H_k\}$; $t(u)=\sum (u-u_0)^k H_k$, where Hamiltonian is $H=H_2$.}

In ABA the {\it Bethe vectors} are special polynomials in the monodromy matrix entries acting on vacuum.  For the
simplest case of the  $\mathfrak{gl}(2)$ related models they are given by\footnote{Let us note,
  that here and further the definition of Bethe vectors $|\bu\rangle$, $|\bu,\bv\rangle$ should not be understood
  as products defined in \eqref{single_prod}. Here $\bu$, $\bv$ denotes just the sets of arguments of
  function.} 
\be{BV}
|\bu\rangle=T_{12}(\bu)|0\rangle.
\ee
For models related to higher rank symmetries we need to apply the so-called {\it nested Bethe
  Ansatz}, where we have different types of Bethe roots corresponding to the different levels of the nesting
procedure.  In concrete physical models the  different types of spectral variables correspond to the
different physical degrees of freedom. In the $\mathfrak{gl}(2|1)$ related models the states are
described  by two types of variables \cite{Kulish85, KULISH1981246}. The explicit form of the Bethe vectors is given by
\be{BV1}
|\bu;\bv\rangle=\sum \frac{g(\bv_{\so},\bu_{\so})f(\bu_{\so},\bu_{\st})g(\bv_{\st},\bv_{\so})h(\bu_{\so},\bu_{\so})}{\lambda_2(\bu)\lambda_2(\bv_{\st})f(\bv,\bu)}\mathbb T_{13}(\bu_{\so})T_{12}(\bu_{\st})\mathbb T_{23}(\bv_{\st})|0\rangle,
\ee
 and the {\it dual Bethe vectors} are given by
\be{dBV1}
\langle\bu;\bv|=(-1)^{b(b-1)}\sum \frac{g(\bv_{\so},\bu_{\so})f(\bu_{\so},\bu_{\st})g(\bv_{\st},\bv_{\so})h(\bu_{\so},\bu_{\so})}{\lambda_2(\bu)\lambda_2(\bv_{\st})f(\bv,\bu)}\langle 0|\mathbb T_{32}(\bv_{\st})T_{21}(\bu_{\st})\mathbb T_{31}(\bu_{\so}).
\ee
In the following we will denote $\#\bu=a$ and $\#\bv=b$. {Above the sum is taken over partitions
  $\bu\to\{\bu_{\so},\bu_{\st}\}$, $\bv\to\{\bv_{\so},\bv_{\st}\}$.} In the general case $\{\bu,
\bv\}$ are sets of arbitrary complex numbers, and we call Bethe vectors 
{\it off-shell}. In the case when the sets $\{\bu,\bv\}$ satisfy the system of {\it Bethe Ansatz
  equations} (BAE) Bethe vectors become eigenvectors {of the transfer matrix}, we call them {\it on-shell}. In our notations
the (twisted) Bethe equations can be written as
\be{BEgl3}
\begin{split}
& r_1(u_j)=\frac{\varkappa_2}{\varkappa_1}\frac{f(u_j,\bu_j)}{f(\bu_j,u_j)}f(\bv,u_j),\qquad j=1,\dots,a,\\
& r_3(v_j)=\frac{\varkappa_2}{\varkappa_3}f(v_j,\bu),\qquad j=1,\dots,b,
\end{split}
\ee
where $\varkappa=\{\varkappa_1,\varkappa_2,\varkappa_3\}$ are {\it twists} (see
\cite{Izergin1984,Korepin1}).

Following \cite{KorepinGaudin} we use the concept of {\it the generalized model}, in which sets $\{r_1(u_k)\}$,   $k=1,\dots,a$,  $\{r_3(v_j)\}$, $j=1,\dots,b$ are treated as sets of free parameters, without any reference to the particular model.  For the off-shell Bethe vector instead of two sets of parameters $\{\bu,\bv\}$ we
have now two additional sets of parameters $\{r_1(u_k)\}$ and $\{r_3(v_j)\}$ that are independent of
$\{\bu,\bv\}$.

The generalized model is a convenient instrument for computations in the ABA approach and will be used throughout the paper. It allows to establish expansions \eqref{LM}--\eqref{LeClairMussardoPozsgayHutsaliuk}  for arbitrary integrable model related to algebra symmetry $\mathfrak{gl}(3)$ and $\mathfrak{gl}(2|1)$.

In sections \ref{corr_total} we specify
functions $\{r_1,r_3\}$ as $r_1=1$, $r_3(v_j)=e^{iv_jL}$ and produce results of
section \ref{results} for the Gaudin-Yang model from the general formulas.

\section{Expansion theorem for the mean values\label{LMseries}}

Here we give a proof LeClair-Mussardo series using standard tools of nested algebraic Bethe Ansatz.

\subsection{Analytic properties of a scalar product}

We define the scalar product $S_{a,b}\left(\buc;\bvc|\bub;\bvb\right)$ of two generic (off-shell) Bethe vectors as
\be{scal}
S_{a,b}\left(\buc;\bvc|\bub;\bvb\right)=\langle \buc;\bvc|\bub;\bvb\rangle.
\ee
Our expansion theorem introduced below is built on certain singularity properties of the scalar
products and form factors. Thus we start with investigating the poles of the scalar product.

It was proven in \cite{SLHRP2} that the scalar product can be expressed via {\it highest
  coefficients} $\mathcal Z_{n,m}$\footnote{Notation $\bu$, $\bv$ should not be confused with
  shorthand notation for products in definitions \eqref{scal}, \eqref{scal21}, $\mathcal
  Z_{a,b}$. These are arguments of function.} 
\be{scal21}
\begin{split}
S_{a,b}\left(\buc;\bvc|\bub;\bvb\right)=\sum \frac{r_1(\buc_{\st})r_1(\bub_{\so})r_3(\bvc_{\st})r_3(\bvb_{\so})}{f(\bvc,\buc)f(\bvb,\bub)}f(\bvc_{\so},\buc_{\so})f(\bvb_{\st},\bub_{\st})\\
\times f(\bub_{\st},\bub_{\so})f(\buc_{\so},\buc_{\st})g(\bvb_{\st},\bvb_{\so})g(\bvc_{\so},\bvc_{\st})\mathcal Z_{a-k,n}\left(\buc_{\st},\bub_{\st}|\bvc_{\so},\bvb_{\so}\right)\mathcal Z_{k,b-n}\left(\bub_{\so},\buc_{\so}|\bvb_{\st},\bvc_{\st}\right).
\end{split}
\ee
Here and further $\#\bub=\#\buc=a$ and $\#\bvb=\#\bvc=b$. The sum is taken over partitions $\bu\to\{\bu_{\so},\bu_{\st}\}$, $\bv\to\{\bv_{\so},\bv_{\st}\}$.

The pole structure of the highest coefficient is known.  {We are interested in residues that appears
  in the limit $\buc\to\bub$, $\bvc\to\bvb$}
\be{HCpole1}
\begin{split}
&\left.\mathcal Z_{a,b}\left(\buc,\bvc|\bub,\bvb\right)\right|_{\uc_a\rightarrow\ub_a}=\frac{ic}{\uc_a-\ub_a}f(\buc_a,\uc_a)f(\uc_a,\bub_a)f(\bvb,\uc_a)\mathcal Z_{a-1,b}\left(\buc_a,\bvc|\bub_a,\bvb\right),\\
&\left.\mathcal Z_{a,b}\left(\buc,\bvc|\bub,\bvb\right)\right|_{\vc_b\rightarrow\vb_b}=\frac{ ic}{\vc_b-\vb_b}g(\vc_b,\bvc_b)g(\vb_b,\bvb_b)f(\vb_b,\buc)\mathcal Z_{a,b-1}\left(\buc,\bvc_b|\bub,\bvb_b\right).
\end{split}
\ee
Let us consider now the analytic properties of the scalar product of two Bethe vectors. If $\uc_a\rightarrow\ub_a$ the r.h.s. of \eqref{scal21} contains poles if $\uc_a\in\buc_{\so}$, $\ub_a\in\bub_{\so}$ or $\uc_a\in\buc_{\st}$, $\ub_a\in\bub_{\st}$. Consider the first case
\be{scal21firstPole1}
\begin{split}
\sum\frac{r_1(\buc_{\st})r_1(\bub_1)r_3(\bvc_{\st})r_3(\bvb_{\so})}{f(\bvc,\buc_a)f(\bvb,\bub_a)}f(\bvc_{\so},\buc_1)f(\bvb_{\st},\bub_{\st})f(\bub_{\st},\bub_{\so})f(\buc_1,\buc_{\st})g(\bvb_{\st},\bvb_{\so})g(\bvc_{\so},\bvc_{\st})\\
\times \frac{ ic r_1(\ub_a)}{\uc_a-\ub_a}\mathcal Z_{a-k,n}\left(\buc_{\st},\bub_{\st}|\bvc_{\so},\bvb_{\so}\right)\mathcal Z_{k-1,b-n}\left(\bub_1,\buc_1|\bvb_{\st},\bvc_{\st}\right)\\
\times f(\bvb_{\st},\uc_a)f(\buc_1,\uc_a)f(\ub_a,\bub_{\so})\frac{f(\bvc_{\so},\uc_a)f(\bub_{\st},\ub_a)f(\uc_a,\buc_{\st})}{f(\bvc,\uc_a)f(\bvb,\ub_a)}.
\end{split}
\ee
Here $\buc_1$, $\bub_1$  denote the sets $\buc_{\so}\setminus \uc_a$ and $\bub_{\so}\setminus \ub_a$. The simplified last line of \eqref{scal21firstPole1} is
\be{simple1}
\begin{split}
f(\buc_a,\uc_a)f(\bub_a,\uc_a) \frac{f(\uc_a,\buc_{\st})}{f(\buc_{\st},\uc_a)}\frac{f(\bub_a,\bub_1)}{f(\bub_1,\ub_a)}\frac{1}{f(\bvc_{\st},\uc_a)f(\bvb_{\so},\ub_a)}.
\end{split}
\ee
The second case is
\be{scal21firstPole2}
\begin{split}
\sum\frac{r_1(\buc_2)r_1(\bub_{\so})r_3(\bvc_{\st})r_3(\bvb_{\so})}{f(\bvc,\buc_a)f(\bvb,\bub_a)}f(\bvc_{\so},\buc_{\so})f(\bvb_{\st},\bub_2)f(\bub_2,\bub_{\so})f(\buc_{\so},\buc_2)g(\bvb_{\st},\bvb_{\so})g(\bvc_{\so},\bvc_{\st})\\
\times \frac{ icr_1(\uc_a)}{\uc_a-\ub_a}\mathcal Z_{a-k-1,n}\left(\buc_2,\bub_2|\bvc_{\so},\bvb_{\so}\right)\mathcal Z_{k,b-n}\left(\bub_{\so},\buc_{\so}|\bvb_{\st},\bvc_{\st}\right)\\
\times \frac{f(\uc_a,\bub_{\so})f(\buc_{\so},\uc_a)f(\bvb_{\st},\ub_a)}{f(\bvc,\uc_a)f(\bvb,\uc_a)}f(\bub_2,\ub_a)f(\uc_a,\buc_2)g(\bvc_{\so},\uc_a)h(\bvc_{\so},\uc_a).
\end{split}
\ee
{Here $\buc_2$, $\bub_2$  denote the sets $\buc_{\st}\setminus\uc_a$, $\bub_{\st}\setminus\ub_a$}.  The simplification of the last line gives
\be{simple2}
f(\buc_a,\uc_a)f(\bub_a,\uc_a) \frac{f(\uc_a,\buc_2)}{f(\buc_2,\uc_a)}\frac{f(\ub_a,\bub_{\so})}{f(\bub_{\so},\ub_a)}\frac{1}{f(\bvc_{\st},\ub_a)f(\bvb_{\so},\ub_a)}.
\ee
Taking the sum of both contributions \eqref{scal21firstPole1} and \eqref{scal21firstPole2} we come
to
\be{renorm_a}
\begin{split}
\left.S_{a,b}\left(\buc;\bvc|\bub;\bvb\right)\right|_{\uc_a\rightarrow\ub_a} = ic\left.\frac{r_1(\uc_a)-r_1(\ub_a)}{\uc_a-\ub_a} f(\buc_a,\uc_a)f(\bub_a,\uc_a) S_{a-1,b}\left(\buc_a;\bvc|\bub_a;\bvb\right)_{mod}\right|_{\uc_a\to\ub_a},
\end{split}
\ee
where in the r.h.s. the vacuum expectation values are modified as
\be{r_mod}
r_1(u)_{mod}=r_1(u)\frac{f(u_a,u)}{f(u,u_a)},\qquad r_3(v)_{mod}=\frac{r_3(v)}{f(v,u_a)}.
\ee
We can take now the limit of first factor
\be{z_lim}
\lim_{\uc_a\to\ub_a}\left(\frac{r_1(\uc_a)-r_1(\ub_a)}{\uc_a-\ub_a}\right)=-iZ_a.
\ee
Consider now the limit $\vc_ b\rightarrow \vb_b$. The first contribution to the residue comes from the partitions where $\vc_a\in\bvc_{\so}$, $\vb_a\in\bvb_{\so}$ and the second from the partition $\vc_a\in\bvc_{\st}$, $\vb_a\in\bvb_{\st}$.  The first contribution is
\be{scal21secondPole1}
\begin{split}
\sum\frac{r_1(\buc_{\st})r_1(\bub_{\so})r_3(\bvc_{\st})r_3(\bvb_1)}{f(\bvc_b,\buc)f(\bvb_b,\bub)}f(\bvc_1,\buc_{\so})f(\bvb_{\st},\bub_{\st})f(\bub_{\st},\bub_{\so})f(\buc_{\so},\buc_{\st})g(\bvb_{\st},\bvb_1)g(\bvc_1,\bvc_{\st})\\
\times \frac{ ic r_3(\vb_b)}{\vc_b-\vb_b}\mathcal Z_{a-k,n-1}\left(\buc_{\st},\bub_{\st}|\bvc_1,\bvb_1\right)\mathcal Z_{k,b-n}\left(\bub_{\so},\buc_{\so}|\bvb_{\st},\bvc_{\st}\right)\\
\times \frac{f(\vc_b,\buc_{\so})g(\bvb_{\st},\vb_b)g(\bvc_{\st},\vc_b)}{f(\vc_b,\buc)f(\vb_b,\bub)}g(\vc_b,\bvc_1)g(\vc_b,\bvb_1)f(\vc_b,\bub_{\st}).
\end{split}
\ee
Here $\bvc_1$, $\bvb_1$  denote the sets $\bvc_{\so}\setminus\vc_b$, $\bvb_{\so}\setminus \vb_b$.
The simplification of the last line gives
\be{simple3}
g(\vc_b,\bvc_b)g(\vc_b,\bvb_b)\frac{1}{f(\vc_b,\buc_{\st})f(\vc_b,\bub_{\so})}.
\ee
The second contribution is
\be{scal21secondPole2}
\begin{split}
\sum\frac{r_1(\buc_{\st})r_1(\bub_{\so})r_3(\bvc_2)r_3(\bvb_{\so})}{f(\bvc_b,\buc)f(\bvb_b,\bub)}f(\bvc_{\so},\buc_{\so})f(\bvb_2,\bub_{\st})f(\bub_{\st},\bub_{\so})f(\buc_{\so},\buc_{\st})g(\bvb_2,\bvb_{\so})g(\bvc_{\so},\bvc_2)\\
\times \frac{icr_3(\vb_b)}{\vc_b-\vb_b}\mathcal Z_{a-k,n}\left(\buc_{\st},\bub_{\st}|\bvc_{\so},\bvb_{\so}\right)\mathcal Z_{k,b-n-1}\left(\bub_{\so},\buc_{\so}|\bvb_2,\bvc_2\right)\\
\times \frac{f(\vc_b,\buc_{\st})g(\bvb_{\so},\vc_b)g(\bvc_{\so},\vb_b)}{f(\vc_b,\buc)f(\vb_b,\bub)}g(\vc_b,\bvc_2)g(\vc_b,\bvb_2)f(\vc_b,\buc_{\so}).
\end{split}
\ee
Here $\bvc_2$, $\bvb_2$  denote the sets $\bvc_{\st}\setminus\vc_b$, $\bvb_{\st}\setminus \vb_b$.
The simplification of the last line gives
\be{contraction_1}
g(\vc_b,\bvc_b)g(\vc_b,\bvc_b)\frac{1}{f(\vc_b,\buc_{\st})f(\vc_b,\bub_{\so})}.
\ee 
Taking the sum of contributions \eqref{scal21secondPole1} and \eqref{scal21secondPole2} we arrive at
\be{Scalp_res}
\left. S_{a,b}\left(\buc;\bvc|\bub;\bvb\right)\right|_{\vc_b\rightarrow\vb_b}
=
\left.  ic\frac{r_3(\vc_b)-r_3(\vb_b)}{\vc_b-\vb_b}g(\vc_b,\bvc_b)g(\vc_b,\bvb_b)S_{a,b-1}\left(\buc;\bvc_b|\bub;\bvb_b\right)_{mod}\right|_{\vc_a\to\vb_a}.
\ee
Here the vacuum expectation values at the r.h.s. are modified as
\be{r1_mod}
r_1(u)_{mod}=r_1(u)/f(v_b,u),\qquad r_3(v)_{mod}=r_3(v),
\ee
and we can take the limit of the first factor there
\be{y_lim}
\lim_{\vc_b\to\vb_b}\left(\frac{r_3(\vc_b)-r_3(\vb_b)}{\vc_b-\vb_b}\right)=-iY_b.
\ee

\subsection{Analytic properties of form factors\label{FF_analytic}}

It is our main goal to study form factors of products of local operators. It is known that in the fundamental models local
operators can be  expressed directly via the entries of the monodromy matrix \cite{Maillet,GohKor}. Therefore, we first study the analytic properties of them.

In models of quantum gases the solution of the quantum inverse problem
is not known. However, taking into  account that such models can be
derived from the fundamental one using a special type of scaling limit
\cite{SGohKlumpSB,GolzerHolz}, we assume that the properties of form
factors in Gaudin-Yang model    are similar to the properties of form
factors in the fundamental model.  This was explicitly
  demonstrated in models related to algebra symmetry
  $\mathfrak{gl}(2)$ in \cite{sajat-XXZ-to-LL}.

We start with the analytic properties of form factors of  the matrix elements of a monodromy matrix. They follow
from the formulas of multiple action of operators $T_{i,j}$ on the Bethe vectors and from the
analytic properties of the scalar product, established earlier. Formulas for the multiple action in the case of 
algebra symmetry $\mathfrak{gl}(2|1)$ were derived in \cite{SLHRP1}.

Consider for instance the multiple action of the operator $T_{31}(\bar z)$
\be{T31}
\begin{split}
\left\langle \buc;\bvc|T_{31}(\bw)|\bub;\bvb \right\rangle=(-1)^{n(n+1)/2}\lambda_2(\bw)h(\bar\xi,\bar z)\sum r_3(\bar\xi_{\so})r_1(\bar\eta_{\st})K_n(\bar\eta_{\st}|\bar\xi_{\st}+c)\\
\times \frac{g(\bar\xi_{\st},\bar\xi_{\so})g(\bar\xi_{\sth},\bar\xi_{\so})g(\bar\xi_{\so},\bar\xi_{\st})}{h(\bar\eta_{\so},\bar z)h(\bar\xi_{\so},\bar\eta_{\so})h(\bar\xi_{\st},\bar z)}\frac{f(\bar\eta_{\so},\bar\eta_{\st})f(\bar\eta_{\so},\bar\eta_{\sth})f(\bar\eta_{\sth},\bar\eta_{\st})h(\bar\eta_{\so},\bar\eta_{\so})}{f(\bar\xi_{\so},\bar\eta_{\st})f(\bar\xi_{\st},\bar\eta_{\sth})f(\bar\xi_{\sth},\bar\eta_{\st})} S_{a,b}\left(\buc;\bvc|\bar\eta_{\sth};\bar\xi_{\sth}\right).
\end{split}
\ee
Here the sum is taken over partition $\{\bu,\bw\}=\bar\eta\rightarrow\{\bar\eta_{\so},\bar\eta_{\st},\dots\}$, $\{\bv,\bw\}=\bar\xi\rightarrow\{\bar\xi_{\so},\bar\xi_{\st},\dots\}$ with the restrictions $\#\bar\xi_{\so}=\#\bar\eta_{\so}=\#\bar z$ and $K_n(\bar x|\bar y)$ is {\it the Izergin-Korepin determinant}, it is defined for any two sets $\bx$, $\by$ with $\#\bx=\#\by=n$:
\be{IK}
K_n(\bx|\by)=\Delta'_g(\bx)\Delta_g(\by)h(\bx,\by)\det_n\left[t(x_j,y_k)\right].
\ee
The first pole in \eqref{T31} is present if $\ub_a\in\bar\eta_{\sth}$. The singular part is proportional to the singular part of products $\langle \buc;\bvc|\bar\eta_{\sth};\bar\xi_{\sth}\rangle$. We write separately only the part under the sum over partitions that depends on $\uc_a$ and $\ub_a$. The following factor appears in front of these singular parts
\be{T31pole1}
\begin{split}
r_1(\bar\eta_{\st})r_3(\bar\xi_{\so})&\frac{f(\bar\eta_{\so},\uc_a)f(\uc_a,\bar\eta_{\st})}{f(\bar\xi_{\so},\uc_a)}f(\buc_a,\uc_a)f(\bar\eta_{\sth},\uc_a)\\
&=f(\bw,\uc_a)f(\buc_a,\uc_a)f(\bub_a,\uc_a)\frac{r_3(\bar\xi_{\so})}{f(\bar\xi_{\so},\uc_a)}r_1(\bar\eta_{\st})\frac{f(\uc_a,\bar\eta_{\st})}{f(\bar\eta_{\st},\uc_a)}.
\end{split}
\ee
After an elementary transformation, we get
\be{T31res1}
\begin{split}
\left.\left\langle \buc;\bvc|T_{31}(\bw)|\bub;\bvb)\right\rangle\right|_{\uc_a\rightarrow\ub_a}
=cZ_a f(\buc_a,\uc_a)f(\bub_a,\uc_a)f(\bw,\uc_a)\left\langle \buc_a;\bvc|T_{31}(\bw)|\bub_a;\bvb\right\rangle_{mod}.
\end{split}
\ee
The second pole is present if $\vb_b\in\bar\xi_{\sth}$. We write only the part under the sum over partition that depends on $\vc_b$ and $\vb_b$
\be{T31pole2}
\begin{split}
g(\vc_b,\bar\xi_{\so})g(\vc_b,\bar\xi_{\st})g(\vc_b,\bvc_b)g(\bvc,\bar\xi_{\sth})\frac{r_3(\bar\eta_{\st})}{f(\vc_b,\bar\eta_{\st})}=g(\vc_b,\bvc_b)g(\vc_b,\bvb_b)g(\vc_b,\bw)\frac{r_3(\bar\eta_{\st})}{f(\vc_b,\bar\eta_{\st})}.
\end{split}
\ee
After an elementary transformation and the replacement $r_1(\bar\eta_{\st})$ by $r_1(\bar\eta_{\st})/f(\vc_b,\bar\eta_{\st})$ we arrive at
\be{T31res2}
\begin{split}
\left.\left\langle \buc;\bvc|T_{31}(\bw)|\bub;\bvb \right\rangle\right|_{\vc_b\rightarrow\vb_b}
=cY_b g(\vc_b,\bvc_b)g(\vc_b,\bvb_b)g(\vc_b,\bw)\left\langle\buc;\bvc_b|T_{31}(\bw)|\bub;\bvb_b\right\rangle_{mod}.
\end{split}
\ee
Since  we have the symmetry w.r.t to permutations of pairs $\{u_k,r_1(u_k)\}\longleftrightarrow \{u_n,r_1(u_n)\}$ for arbitrary $k, n=1,\dots,a$ and $\{v_k,r_3(v_k)\}\longleftrightarrow \{v_n,r_3(v_n)\}$ for arbitrary $k, n=1,\dots,b$ the same property is true for each rapidity in the set $\bu$ and $\bv$.

Using the action rules for all $T_{ij}(\bw)$, $i,j=1,\dots,3$ it is easy to obtain that the same
analytical properties will hold for arbitrary $T_{ij}(\bw)$. It is also easy to check in the same
way that the same analytic properties holds for superposition of operators
$T_{ij}(\bw_{\so})T_{kl}(\bw_{\st})\dots$. Thus, the form factor of an arbitrary operator $O$ will
satisfy the same properties:
\be{Ores1}
\begin{split}
\left.\left\langle \buc;\bvc|O(\bw)|\bub;\bvb)\right\rangle\right|_{\uc_a\rightarrow\ub_a}
=c{ r_1(\bub_a)}Z_a f(\buc_a,\uc_a)f(\bub_a,\uc_a)f(\bw,\uc_a)\left\langle \buc_a;\bvc|O(\bw)|\bub_a;\bvb\right\rangle_{mod},
\end{split}
\ee
\be{Ores2}
\begin{split}
\left.\left\langle \buc;\bvc|O(\bw)|\bub;\bvb \right\rangle\right|_{\vc_b\rightarrow\vb_b}
=c{ r_3(\bvb_b)}Y_b g(\vc_b,\bvc_b)g(\vc_b,\bvb_b)g(\vc_b,\bw)\left\langle\buc;\bvc_b|O(\bw)|\bub;\bvb_b\right\rangle_{mod}.
\end{split}
\ee
The only difference in the case  of algebra symmetry $\mathfrak{gl}(3)$ will be replacement of the
factor $g(\vc_b,\bvc_b)g(\vc_b,\bvb_b)g(\vc_b,\bw)$ in the last line by
$f(\vc_b,\bvc_b)f(\vc_b,\bvb_b)f(\vc_b,\bw)$.
  
\subsection{Expansion formula\label{expansion}}

The proof of the expansion formula is loosely based on the proof for the algebra symmetry
$\mathfrak{gl}(2)$ case given in \cite{Pozsgay2,Pozsgay1}. Consider the diagonal form factor
\be{diagonal_definition}
F_{a,b}^O\left(\bu;\bv|\bZ;\bY\right)=
\frac{1}{f(\bw,\buc)g(\bvc,\bw)}\langle \buc;\bvc|O|\bub;\bvb\rangle|_{\buc\to\bub,\:\bvc\to\bvb}.
\ee
Here it is understood that the diagonal limit is performed on the off-shell form factor.
It follows from the above that the diagonal form factor is a linear function of the variables $Z_j$,
$Y_k$ with the coefficients $cf(\buc_a,\uc_a)f(\bub_a,\uc_a)$ and
$cg(\vc_b,\bvc_b)g(\vc_b,\bvb_b)$\footnote{Factors  $f(\bw,\uc_a)$ and $g(\vc_b,\bw)$ are absorbed
  here into the normalization of the form factor (see \eqref{diagonal_definition}).}. Taking the
limit of $\buc\to \bub$, $\bvc\to\bvb$ in \eqref{Ores1}, \eqref{Ores2} we arrive at
\be{Recursion}
\begin{split}
&\frac{\partial}{\partial Z_a}F_{a,b}^O\left(\bu;\bv|\bZ;\bY\right)=c{ r_1(\bub_a)}f(\buc_a,\uc_a)f(\bub_a,\uc_a)F_{a-1,b}^O\left(\bu_a;\bv|\bZ_a;\bY\right)_{mod},\\
&\frac{\partial}{\partial Y_b}F_{a,b}^O\left(\bu;\bv|\bZ;\bY\right)=c{r_3(\bvb_b)}g(\vc_b,\bvc_b)g(\vc_b,\bvb_b)F_{a,b-1}^O\left(\bu;\bv_b|\bZ;\bY_b\right)_{mod}.
\end{split}
\ee
Now we introduce {\it the symmetric form factor}. As given in the previous Section, its definition
in the general case is
\be{sym_limit}
\mathbb  F_{a,b}\left(\bu;\bv\right)=
\lim_{\varepsilon\to 0}\langle \bub+\varepsilon;\bvb+\varepsilon|O|\bub;\bvb\rangle.
\ee
Here we regard $\{\buc,\bvc\}$ and $\{\bub,\bvb\}$ as solutions of the Bethe equations, i.e. limit
$\buc\to\bub$, $\bvc\to\bvb$ is taken after the  $r_1(u_k)$, $r_3(v_k)$ are expressed via r.h.s. of
\eqref{BEgl3} (without twists). 

 In the case of models related to symmetry $\mathfrak{gl}(2)$ it was proven already in 
\cite{Pozsgay1} that the symmetric form factor is the mean value in the case when the
parameters $\bar Z$ are chosen to be zero. In the nested case a similar property holds:
\be{FSFO00}
\mathbb  F_{a,b}\left(\bu;\bv\right)\equiv
F_{a,b}^O\left(\bu;\bv|\{0\};\{0\}\right).
\ee

\begin{thm}

{For the (in general non-local) operator $O$ and arbitrary eigenstates the following expansion of form factor can be established}
\be{expand}
F_{a,b}^O\left(\bu;\bv|\bZ;\bY\right)=\sum S_{m, n}(\bu_{\so};\bv_{\so}|\bZ_{\so};\bY_{\so})\mathbb  F_{a-m,b-n}\left(\bu_{\st};\bv_{\st}\right),
\ee
where 
\be{S_def}
\begin{split}
S_{a, b}\left(\bu_{\so};\bv_{\so}|\bar Z_{\so};\bar Y_{\so}\right)=f(\bv_{\so},\bu)f(\bv_{\st},\bu_{\so})\Delta_f(\bu_{\so})\Delta_f'(\bu_{\so})\Delta_g(\bv_{\so})\Delta_g'(\bv_{\so})\\
\times f(\bu_{\so},\bu_{\st})f(\bu_{\st},\bu_{\so})g(\bv_{\st},\bv_{\so})g(\bv_{\so},\bv_{\st})\det_{a+b} G\left(\bu_{\so},\bv_{\so}\right),
\end{split}
\ee
and $G$ is {\it the Gaudin determinant}. The cardinalities of the sets are $\#\bu_{\so}=m$, $\#\bv_{\so}=n$ and $\#\bu_{\st}=a-m$, $\#\bv_{\st}=b-n$.
\end{thm}
Proof. Formula \eqref{expand} can be proven by induction. Let us start from the first nontrivial case $\#\bu_{\st}\equiv a_2=a$, $\#\bv_{\st}\equiv b_2=b$ for which  there is one non-zero term in the r.h.s. of \eqref{expand} and this is $\mathbb F_{a,b}$. This term is $\bZ$- and $\bY$-independent.  The l.h.s. is, of course, $\bZ$-, $\bY$-independent too, it is clear from \eqref{Recursion}. 

The matrix elements of the Gaudin matrix are given later, see \eqref{P11-n2}--\eqref{P12-n2}. It is clear from that explicit form that the derivatives of $S_{a, b}$ w.r.t. $Z_j$, $Y_k$ are
\be{Sfunc}
\begin{split}
&\frac{\partial}{\partial Z_j}S_{m,n}\left(\bu;\bv|\bar Z;\bar Y\right)=c f(\bv,u_j)f(u_j,\bu_j)f(\bu_j,u_j)S_{m-1,n}\left(\bu_j;\bv|\bZ_j;\bY\right)_{mod},\\
&\frac{\partial}{\partial Y_j}S_{m,n}\left(\bu;\bv|\bar Z;\bar Y\right)=c f(v_j,\bu)g(v_j,\bv_j)g(\bv_j,v_j)S_{m,n-1}\left(\bu;\bv_j|\bZ;\bY_j\right)_{mod},
\end{split}
\ee
here
{$Z_k$, $Y_j$ are modified as
\be{z1mod}
\begin{split}
&cZ'_j=cZ_j-K(u_j,u_a),\\
&cY'_j=cY_j-t(v_j,u_a),
\end{split}
\ee
in first case and as
\be{z2mod}
\begin{split}
&cZ'_j=cZ_j+t(v_b,u_j),\\
&cY'_j=cY_j,
\end{split}
\ee
in the second case.
}

Supposing that \eqref{expand} holds for some arbitrary $a-1$ we can prove that it holds for $a$
also. Take the derivative of \eqref{expand} w.r.t. $Z_j$. Using \eqref{Sfunc} in the r.h.s. of
\eqref{expand} we immediately reproduce the same expansion but with set $\bu$ replaced by $\bu_a$
that is, according to the induction assumption, $F^O(\bu_a;\bv|\bZ_a; \bY)$. At the l.h.s. we use
\eqref{Recursion} and obtain exactly the same. Therefore the $\bZ$-dependent part coincides. The same
procedure can be done for $\bY$-dependent part. In order to compare $\bZ$- and $\bY$-independent
part it is enough to send all $\bZ$ and $\bY$ to zero. In this case only one contribution $\mathbb
F_{a,b}(\bu;\bv)$ survives and it is equal to $F^O_{a,b}\left(\bu;\bv|\{0\};\{0\}\right)$ in the
l.h.s. by definition of $\mathbb  F_{a,b}(\bu;\bv)$. \qed

After dividing \eqref{expand} by norm of eigenvectors $\langle\bu;\bv|\bu;\bv\rangle$ (see \cite{SLHRP3}) we reproduce formula \eqref{LM}
\be{LeClairMussardo}
\langle O\rangle_{a,b}\left(\bu;\bv\right)=\frac{\sum\det G\left(\bu_{\so},\bv_{\so}\right) \mathfrak F_{a-m,b-n}\left(\bu_{\st};\bv_{\st}\right)}{\det G\left(\bu,\bv\right)}.
\ee
Here $\langle O\rangle$ in the l.h.s. denotes the normalized expectation value and $\mathfrak F_{ m,n}$ denotes normalized $\mathbb  F_{m,n}$
\be{F_norm}
\mathfrak F_{m,n}\left(\bu;\bv\right)=\frac{\mathbb F_{S\; m,n}(\bu;\bv)}{f(\bv,\bu)\Delta_f(\bu)\Delta'_f(\bu)\Delta_g(\bv)\Delta'_g(\bv)}.
\ee
{It is easy to prove, that  formula \eqref{LeClairMussardo} 
is still valid in the case of algebra symmetry $\mathfrak{gl}(3)$ up to replacement of the Gaudin
matrix $G(\bu;\bv)$ and $\mathfrak F_{ a,b}$ by the corresponding analogs.}

\subsection{Correlation functions via the connected form factors\label{sec:connected}}

The original LeClair-Mussardo series \cite{LM} was built on the so-called connected form factors, which were defined as the pole free part of the diagonal limit of the on shell form factors. The relations between the symmetric and connected form factors were later clarified in \cite{Takacs1}. In this work we present the LM series using the connected form factors also in the
nested {Bethe Ansatz}.

The connected form factors can be defined similarly to \cite{LM,Takacs1} through the limit
\be{connected_limit}
\mathfrak F^c_{a,b}(\bu;\bv)=\Fin\left\{\lim_{\bar\varepsilon\to0,\;\bar\varepsilon'\to0}\langle \bub+\bar\varepsilon;\bvb+\bar\varepsilon'|O|\bub;\bvb\rangle\right\},
\ee
where the left Bethe vectors is defined as
\be{left_BV}
\langle\bub+\bar\varepsilon;\bvb+\bar\varepsilon'|=\langle\ub_1+\varepsilon_1,\dots, \ub_a+\varepsilon_a;\vb_1+\varepsilon'_1,\dots, \vb_b+\varepsilon'_b|,
\ee
and finite part $\Fin\{\dots\}$ is defined in such a way that all singular terms including ratios of
form $\varepsilon_i/\varepsilon_j$, $\varepsilon'_i/\varepsilon'_j$ for $i\ne j$ and
$\varepsilon'_i/\varepsilon_j$, $\varepsilon_i/\varepsilon'_j$ are discarded.

Combining the arguments of \cite{Takacs1,Pozsgay2} 
and our computations with the scalar products and form factors in the
nested case it is possible to prove the following expansion.

{  
\begin{thm} For the diagonal form factor of the (in general non-local) operator $O$ between the arbitrary eigenvectors the following expansion holds
\be{LeClairMussardo_connected}
\langle O\rangle_{a,b}\left(\bu;\bv\right)=\frac{\sum\det\bar G\left(\bu_{\so},\bv_{\so}|\bu;\bv\right) \mathfrak F^c_{a-m,b-n}\left(\bu_{\st};\bv_{\st}\right)}{\det G\left(\bu,\bv\right)}.
\ee
Here  $\bar G\left(\bu_{\so},\bv_{\so}|\bu;\bv\right)$ is defined as the
matrix built from the columns and rows of the Gaudin matrix
$G(\bu;\bv)$ that belong to the subsets $\bu_{\so}$, $\bv_{\so}$, and
we stress explicitly that in contrary to $G(\bu_{\so};\bv_{\so})$ this
matrix  depends on the full sets of parameters $\bu$, $\bv$. The
normalization of $\mathfrak F^c$ here coincides with the normalization
of \eqref{F_norm}. 
\end{thm}
}

\section{Thermodynamic limit\label{thermodynamic_limit}}

In this Section we take the thermodynamic limit of the expansion theorem, leading to the
LM series in nested Bethe Ansatz systems.

\subsection{Distribution of Bethe roots}

The distribution of Bethe parameters in the thermodynamic limit is given by integral equations
for densities, which follow from the Bethe equations
\cite{TakahashiGas}. {We restrict ourselves to cases where, at
  least in the ground state, solutions of the BAE are real;
the Gaudin-Yang model with $c>0$ belongs to this class of models.
  In such a case  the integral equations for the ground state are}
\be{particles_holes}
\begin{split}    
\rho_1(\lambda)&=\frac{p_1'(\lambda)}{2\pi}-\int_{-B}^B\frac{d\omega}{2\pi c}\:K(\lambda,\omega)\rho_1(\omega)+
\int_{-Q}^Q\frac{d\omega}{2\pi c}\:t(\lambda,\omega)\rho_2(\omega),\\
\rho_2(\lambda)&=\frac{p_2'(\lambda)}{2\pi c}+\int_{-B}^B \frac{d\omega}{2\pi c}\:t(\lambda,\omega)\rho_1(\omega),
\end{split}
\ee
where $\rho_1$, $\rho_2$ are densities of particles, $p'_1(x)=-i\partial \log r_1(x)$,
$p'_2(x)=-i\partial \log r_3(x)$ and $B$ (resp. $Q$ are Fermi boundaries for the first (resp. second)
level of Bethe rapidities.

In the Gaudin-Yang model total densities $n_{\up/\down}$ depend on $Q$ and $B$, and in principle this relation can be
inverted to find $Q$ and $B$ in terms of the physical quantities $n_{\up/\down}$. In
practice this involves the solution of the linear integral equations above. The relations simplify
in the large imbalance limit, as discussed in detail in Section \ref{densities}.

\subsection{Ratio of Gaudin determinants: Fermi case \label{Gaudin_thermodynamic_fermi}}

Here we calculate the ratios of the Gaudin determinants $\det G(\bu_{\so},\bv_{\so})$ and $\det
G(\bu,\bv)$ in the thermodynamic limit. We use the approach developed earlier \cite{Korepin1,
  Izergin1984, Pozsgay2}. The Gaudin determinant for the algebra symmetry $\mathfrak{gl}(2|1)$
related models was calculated in \cite{SLHRP3}.
\be{Norm-res}
\langle \bu;\bv|\bu;\bv\rangle=(-1)^{a+b}f(\bv,\bu)\Delta_f(\bu)\Delta'_f(\bu)\;
\Delta_g(\bv)\Delta'_g(\bv)\;\det_{a+b}G(\bu,\bv).
\ee
$G$ is an $(a+b)\times(a+b)$ block-matrix. The diagonal blocks are
\be{P11-n2}
\begin{split}
&G_{j,k}=
\delta_{jk}\left[c\,Z_k-\sum_{\ell=1}^aK(u_k,u_{\ell})-\sum_{m=1}^b t(v_m,u_k)  \right]+
K(u_k,u_j),\qquad j,k=1,\dots,a,\\
&G_{j+a,k+a}=\delta_{jk}\left[c\,Y_k+\sum_{\ell=1}^a t(v_k,u_\ell) \right],\qquad j,k=1,\dots,b.
\end{split}
\ee
The anti-diagonal blocks are
\be{P12-n2}
\begin{split}
&G_{j,k+a}=t(v_k,u_j), \qquad j=1,\dots,a,\qquad k=1,\dots,b,\\
&G_{j+a,k}=-t(v_j,u_k),\qquad j=1,\dots,b,\qquad k=1,\dots,a.
\end{split}
\ee
Let us denote
\be{theta-def}
\theta^{(1)}_k=c\,Z_k-\sum_{\ell=1}^aK(u_k,u_{\ell})-\sum_{m=1}^b t(v_m,u_k),\qquad
\theta^{(2)}_{k+a}=c\,Y_k+\sum_{\ell=1}^a t(v_k,u_\ell). 
\ee
After extracting these diagonal parts from the Gaudin determinant, it can be presented as a product
\be{Gaudin_decompos}
\det G=\det \theta\det\widehat{G},
\ee
\be{pre_Fredholm}
\widehat{G}=
\left(\begin{BMAT}(@,70pt,45pt){c:c}{c:c}
\delta_{ij}+\frac{K(u_j,u_k)}{L\theta^{(1)}(u_j)} & \frac{t(v_k,u_j)}{L\theta^{(1)}(u_j)}\\
-\frac{t(v_j,u_k)}{L\theta^{(2)}(v_j)} &  \delta_{ij}
\end{BMAT}\right),
\ee
and $\det\theta$ is a determinant of a diagonal matrix with components $\{\theta^{(1)}_1,\dots,\theta^{(1)}_a,\theta^{(2)}_{a+1},\dots,\theta^{(2)}_{a+b}\}$. 

Consider first the case $\bu_{\so}=\bu_j$ ,  $\bv_{\so}=\bv$ (or $\bu_{\so}=\bu$,  $\bv_{\so}=\bv_j$). Denote as $\det\theta_j'$ the determinant extracted from $\det G(\bu_j,\bv)$ (or from $\det G(\bu,\bv_j)$).  The ratios of $\det\theta'_j/\theta$ for the first and the second case can be presented as
\be{ratio_Gaudin}
\begin{split}
&\frac{\det\theta'_j}{\det\theta}=\frac{1}{\theta^{(1)}(u_j)}\prod_{\ell=1;\ell\ne j}^a\left(\frac{\theta^{(1)}(u_{\ell})+K(u_j,u_{\ell})}{\theta^{(1)}(u_{\ell})}\right)\prod_{\ell=1}^b\left(\frac{\theta^{(2)}(v_{\ell})-t(v_{\ell},u_j)}{\theta^{(2)}(v_{\ell})}\right), \qquad j=1,\dots,a,\\
&\frac{\det\theta'_j}{\det\theta}=\frac{1}{\theta^{(2)}(v_j)}\prod_{\ell=1}^a\left(\frac{\theta^{(1)}(u_{\ell})+t(v_j,u_{\ell})}{\theta^{(1)}(u_{\ell})}\right), \qquad j=1,\dots,b.
\end{split}
\ee
In the thermodynamic limit
\be{thermo_density}
\theta^{(1)}(u_j)=-c L\rho^{(1)}(u_j),\qquad \theta^{(2)}(v_j)=-c L\rho^{(2)}(v_j),
\ee
and \eqref{ratio_Gaudin} correspondingly leads to
\be{thermo_ratio_Gaudin}
\begin{split}
&\frac{\det\theta'_j}{\det\theta}\equiv\frac{1}{L\rho^{(1)}(u_j)}\omega^{(1)}(u_j)
\qquad j=1,\dots,a,\\
&\frac{\det\theta'_j}{\det\theta}\equiv\frac{1}{L\rho^{(2)}(v_j)}\omega^{(2)}(v_j)
\qquad j=1,\dots,b,
\end{split}
\ee
where we defined the weight functions
\be{omega}
\begin{split}
\omega^{(1)}(u_j)&={\frac{1}{c}}\exp\left(-\frac{1}{2c\pi}\int_{-B}^B du\:K(u,u_j)+\frac{1}{2c\pi}\int_{-Q}^Q dv\: t(v,u_j)\right),
\qquad j=1,\dots,a,\\
\omega^{(2)}(v_j)&=\frac{1}{c}\exp\left(-\frac{1}{2c\pi}\int_{-B}^B du\:t(v_j,u)\right), \qquad j=1,\dots,b.
\end{split}
\ee
{Note that these weights contain an explicit factor of $c^{-1}$; this factor comes from our
  choice of normalization of the Bethe states and thus the form factors. This factor is not included
  in the $\omega$-function found in the works \cite{Izergin1984,
    Korepin1,KorepinSlavnov,Izergin1985}; instead it is carried by their form factors.}

In the thermodynamic limit  the remaining part of  \eqref{Gaudin_decompos}, $\det\hat G(\bu,\bv)$
leads to a Fredholm determinant with the kernel $K$, similarly to the algebra symmetry $\mathfrak{gl}(2)$ related models \cite{Korepin1}:
\be{det_G_hat}
\det\hat G=\det\left(\mathbb I-\widehat{K}\right),
\ee
\be{G_hat}
\widehat{K}
\begin{bmatrix}
F^{(1)}(u)\\
F^{(2)}(v)
\end{bmatrix}\equiv
\frac{1}{c}
\left(\begin{BMAT}(@,70pt,45pt){c:c}{c:c}
-\int_{-B}^B du\; K(u,u_k)F^{(1)}(u) & \int_{-Q}^Q du\; t(v_k,u)F^{(1)}(u)\\
-\int_{-B}^B dv\; t(v,u_k)F^{(2)}(v)&  0
\end{BMAT}\right).
\ee
Since $\det\hat G(\bu_j,\bv)$ and $\det\hat G(\bu,\bv_j)$ become the same Fredholm determinant in the thermodynamic limit,
\be{thermo_ratio_Fredholm}
\det\hat G(\bu_j,\bv)/\det\hat G(\bu,\bv)=\det\hat G(\bu,\bv_j)/\hat G(\bu,\bv)=1+O(L^{-1}). 
\ee
In the case $\bu_{\so}=\bu_{j_1,\dots,j_k}$ (or $\bv_{\so}=\bv_{j_1,\dots,j_k}$) the same recipe can be
applied with minor changes. Each $u_{j_1},\dots, u_{j_k}$ (and each $v_{j_1},\dots,v_{j_k}$) will
produce a corresponding density function in the denominators of  \eqref{thermo_ratio_Gaudin} and the
corresponding weight in the exponent. The approach remains valid as long as we restrain ourselves to
$\#\bv_{\so}\sim \#\bv$ and $\#\bu_{\so}\sim\#\bu$.
In practice we always truncate the series \eqref{LeClairMussardo} to small values of {
  $\#\bu_{\st}$, $\#\bv_{\st}$},
so this restriction is always respected.
  
\subsection{Thermodynamic limit of the expansion formula\label{expansion_thermo}}

Now we are able to take thermodynamic limit of \eqref{LeClairMussardo}. Repeating the steps from
\cite{Pozsgay2} and taking into account \eqref{Gaudin_decompos}, \eqref{thermo_ratio_Gaudin},
\eqref{thermo_ratio_Fredholm} we arrive at the following expression 
\be{LeClairMussardoPozsgay}
\langle O\rangle=\sum_{m,n=0}^{\infty}\frac{1}{m!n!}\int_{-B}^B\frac{d\bar t}{(2\pi)^m} \int_{-Q}^Q \frac{d\bar s}{(2\pi)^n}\;\mathfrak\omega^{(1)}(\bar t)\omega^{(2)}(\bar s) \mathfrak F_{m,n}\left(\bar t;\bar s\right),
\ee
where $\#\bar t=m$ and $\#\bar s=n$. 
{This is one of our main results; it can be considered as a
  generalization of the LeClair-Mussardo formula to nested Bethe
  Ansatz systems.}
The computation of the symmetric form factors entering the series is
presented in Section \ref{correlation_functions}. 

It can be shown that formally the same series also holds in the case of algebra symmetry
$\mathfrak{gl}(3)$. The only difference is the replacement of the functions $\omega^{(2)}(\bv)$ by
\be{omega3}
\omega^{(2)}(v_j)={\frac{1}{c}}\exp\left(\frac{1}{2c\pi}\int_{-Q}^Q dv\:K(v,v_j)-\frac{1}{2c\pi}\int_{-B}^B du\: t(v_j,u)\right),
\qquad j=1,\dots,b,
\ee
under the integral
and in the explicit form of the coefficients $\mathfrak F_{m,n}$.  

\subsection{Thermodynamic limit of the connected form factor series\label{expansion_c_thermo}}

{It can be shown absolutely similarly to a case of algebra symmetry
$\mathfrak{gl}(2)$ related models \cite{Pozsgay2} that the
thermodynamic limit of the expansion theorem using the connected form factors is given by}
\be{LeClairMussardoPozsgayc2}
\langle O\rangle=\sum_{m,n=0}^{\infty}\frac{1}{m!n!}\int_{-B}^B\frac{d\bar t}{(2\pi c)^m}
\int_{-Q}^Q \frac{d\bar s}{(2\pi c)^n}\; \mathfrak F^c_{m,n}\left(\bar t;\bar s\right).
\ee

\section{Correlation functions\label{correlation_functions}}

The techniques developed above allow to expand correlation functions
in a series of  form factors (symmetric or connected). The
representation is given by an infinite integral series, and it is a
natural question whether it is possible to obtain  meaningful
expressions for correlators via this series.  

In the case of the Lieb-Linger gas
   it was observed that the series
similar to \eqref{LeClairMussardoPozsgay} has very good convergence in
the strong coupling limit \cite{Izergin1984,Korepin1,Pozsgay2}. This was
also justified from the field theory limit and checked numerically in
\cite{KormosMussardoTrombettoni1,KormosMussardoTrombettoni2} for the
case of correlator $(\psi^{\dagger})^k(\psi)^k$, $k=2,3$.  
Furthermore, the expansion of correlation functions using relatively similar to our approach studied by \cite{Korepin1, Izergin1985} also possess good convergence.

We expect that the similar properties will hold in the two-component models. We show below, that in the
spin-1/2 fermion model the convergence is guaranteed if the two dimensionless parameters $Q/c$
and $B/c$ are small. Smallness of the two combinations corresponds to
the strong coupling and strong imbalance between the components with
different spin projections.

\subsection{Generating functions\label{gen_f}}

As it was pointed in Section \ref{FF_analytic} we expect that the LM-series can be applied also in the non-fundamental models, despite that it is often not easy to express physical operators via the monodromy matrix entries in the general case. Often it can be done by proper scaling limit from the fundamental model; here, however, we use an approach that allows to compute the two-point correlation functions in a more direct way.

We denote by $L$ the system size, the segment $[0,x]$ will be referred as the first subsystem while
$[x,L]$ will be the second. Consider the generating function for the
form factors of diagonal operators  $\exp\left(\alpha Q^{(1)}\right)$, here $\alpha
\Qone=\alpha_1\Qone_1+\alpha_2\Qone_2$, where $\Qone_k$, $k=1,2$ are operators
that measure the numbers of particles of type $k$ in the subsystem 1
and $\alpha_1$, $\alpha_2$ we call {\it twist parameters}. The
generating function $\exp\left(\alpha Q_1\right)$ contains enough information so
that it yields both the  one-point and two-point correlators. It is
proven in Appendix \ref{quasiReshet}, that using the concept of the
partial model (see description after \eqref{lambda}), similarly to the
case of the algebra symmetry $\mathfrak{gl}(2)$ related models, the
(un-normalized) form factor of $O=\exp\left(\alpha \Qone\right)$ can be presented as 
\be{sum_form_final}
\begin{split}
\langle\buc;\bvc|O|\bub;\bvb\rangle=\langle \buc;\bvc |\exp\left(\alpha \Qone\right) |\bub;\bvb \rangle&=\sum \frac{\ell_1(\buc_{\st})\ell_3(\bvb_{\so})}{\ell_1(\bub_{\st})\ell_3(\bvc_{\so})}g(\bvc_{\st},\bvc_{\so})g(\bvb_{\so},\bvb_{\st})\\
\times f(\bvc_{\so},\buc_{\so})f(\bvb_{\st},\bub_{\st})f(\buc_{\so},\buc_{\st})f(\bub_{\st},\bub_{\so})&\Theta^{\alpha}_{a_1, b_2}\left(\bub_{\so};\buc_{\so}|\bvb_{\st};\bvc_{\st}\right)\Theta^{\alpha}_{a_2, b_1}\left(\bub_{\st};\buc_{\st}|\bvb_{\so};\bvc_{\so}\right),
\end{split}
\ee
where it is understood that
 $\{\bub,\bvb\}$ and $\{\buc,\bvc\}$ satisfy BAE \eqref{BEgl3} with a trivial twist
 $\varkappa_1=\varkappa_2=\varkappa_3=1$. In the formula above
$\Theta^{\alpha}_{m,n}$ is a quantity that will be called {\it the highest
  coefficient}  following \cite{Izergin1984,Korepin1}\footnote{Not to be confused with the highest
  coefficient $\mathcal Z_{a,b}$ of the scalar product 
of two Bethe vectors used in earlier section \ref{LMseries}. We use
the same expression due to historical reasons.}.
It is proven in Appendix \ref{quasiReshet} that it coincides with a scalar product of the on-shell and the twisted-on-shell Bethe vectors with a twist $\alpha=\{\alpha_1,0,\alpha_2\}$:
\be{Theta}
\Theta^{\alpha}_{a, b}(\bub;\buc|\bvb;\bvc)=\langle\buc;\bvc|\bub;\bvb\rangle, \qquad \#\bub=\#\buc=a,\qquad \#\bvb=\#\bvc=b.
\ee
Here $\ell_i=r^{(1)}_i$, $i=1,3$  (partial $r_i$, that describe the first subsystem, see the description  above  \eqref{zy}). 

Correlation functions of densities  can be expressed via the generating function using the generation functions.  Thus in the case of Gaudin-Yang model
{
\be{Q_def}
\Qone_i=\int_{0}^x dz\; q_i(z),\qquad \Qtwo_i=\int_{x}^L dz\; q_i(z),
\ee
}
\be{Gen_func_gl_21}
\langle q(x)q(0)\rangle=-\frac{1}{2}\left.\frac{\partial^2}{\partial x^2}\frac{\partial^2}{\partial\alpha^2_2}\left\langle\exp\left(\alpha \Qone\right)\right\rangle\right|_{\alpha=0},
\ee
and
\be{Gen_func_gl_21_2}
\langle q(x)q_{\downarrow}(0)\rangle=\left.\frac{\partial^2}{\partial x^2}\frac{\partial^2}{\partial\alpha_1\partial\alpha_2}\left\langle\exp\left(\alpha_1 \Qone_1+\alpha_2\Qtwo_2\right)\right\rangle\right|_{\alpha=0}
=-\left.\frac{\partial^2}{\partial x^2}\frac{\partial^2}{\partial\alpha_1\partial\alpha_2}\left\langle e^{\alpha_2 b}\exp\left(\alpha \Qone\right)\right\rangle\right|_{\alpha=0}.
\ee
Here $\Qtwo_2=b-\Qone_2$ is a particle number in the second subsystem $[x,L]$, and in the last formula we replace $\alpha_2\to-\alpha_2$ that just changes the common sign. Notation $\alpha=0$ here and further means that $\alpha_1=\alpha_2=0$.

In a similar way correlators in other models can be expressed via derivatives of $\langle \exp(\alpha \Qone)\rangle$ (for example correlators of electrons densities in super-symmetric t-J model, densities in the Fermi-Bose mixtures). For the lattice models the derivatives w.r.t. $x$ would be naturally replaced by the finite differences.

Taking the second derivative of \eqref{sum_form_final} w.r.t. $\alpha_s$, $s=1,2$ at $\alpha=0$ and taking into account, that $\Theta_{m,n}^{\alpha=0}(\bub;\buc|\bvb;\bvc)=\delta_{m,0}\delta_{n,0}$  due to the orthogonality of the on-shell Bethe vectors, we arrive at
\be{deriv_1}
\begin{split}
\left.\frac{\partial^2}{\partial\alpha_i\partial\alpha_j}\langle\buc;\bvc|O|\bub;\bvb\rangle\right|_{\alpha=0}=2\sum_{a_1,b_1}&\frac{\ell_1(\buc_{\st})\ell_3(\bvb_{\so})}{\ell_1(\bub_{\st})\ell_3(\bvc_{\so})}g(\bvc_{\st},\bvc_{\so})g(\bvb_{\so},\bvb_{\st})\\
\times f(\bvc_{\so},\buc_{\so})f(\bvb_{\st},\bub_{\st})f(\buc_{\so},\buc_{\st})f(\bub_{\st},\bub_{\so})&\frac{\partial}{\partial\alpha_i}\Theta_{a_1, b_2}\left(\bub_{\so};\buc_{\so}|\bvb_{\st};\bvc_{\st}\right)
\frac{\partial}{\partial\alpha_j}\Theta_{a_2, b_1}\left(\bub_{\st};\buc_{\st}|\bvb_{\so};\bvc_{\so}\right)\\
+\frac{\ell_1(\buc)\ell_3(\bvb)}{\ell_1(\bub)\ell_3(\bvc)}\frac{\partial^2}{\partial\alpha_i\partial\alpha_j}&\Theta_{a,b}\left(\buc;\bub|\bvc;\bvb\right)+\frac{\partial^2}{\partial\alpha_i\partial\alpha_j}\Theta_{a,b}\left(\buc;\bub|\bvc;\bvb\right),
\end{split}
\ee
here the derivatives are taken at $\alpha=0$, we omit superscript $\alpha$ in this case, and the
operator in question is $O=\exp(\alpha \Qone)$. The terms with $\#\buc_{\st}=\#\bub_{\st}=a_2=a$,
$\#\bvc_{\so}=\#\bvb_{\so}=b_1=b$ and $\#\buc_{\so}=\#\bub_{\so}=a_1=a$, $\#\bvc_{\st}=\#\bvb_{\st}=b_2=b$ are written separately and  $0\le a_1\le a$, $0\le b_1\le b$ with the restriction that the summation over
  $\{a_1, b_1\}$ excludes the cases $\{0,0\}$ and $\{a,b\}$.
\begin{prop}\label{trivial}

If  $\ell_1(\uc)=1$, $\ell_1(\ub)=1$, $\ell_3(\vc)=1$, $\ell_3(\vb)=1$ then
\be{trivial_model}
\left.\frac{\partial^2}{\partial\alpha_i\partial\alpha_j}\langle\buc;\bvc|O|\bub;\bvb\rangle\right|_{\alpha=0}=0.
\ee
\end{prop}

{\it Proof. } Following the arguments of \cite{Izergin1984,Korepin1} we consider a special model in which the monodromy matrix of the first subsystem is given by the unit matrix
\be{spec_monodromy}
T_1(u)=\diag\{1,1,1\}.
\ee
In this model $\ell_1=1$ and $\ell_3=1$. Since $T_{12}(u)$, $T_{13}(u)$ and $T_{23}(u)$ are equal to zero and $\left(Q_1^kQ_1^j\right) |\bu;\bv\rangle=0$ for $k,j=1,2$, we realize that $\partial_{\alpha_i}\partial_{\alpha_j}\langle O\rangle_{a,b}=0$ in the case $\ell_1=\ell_3=1$.
\qed
As an obvious consequence of the proposition \ref{trivial} we can rewrite \eqref{deriv_1} as 
\be{derive_2}
\begin{split}
\left.\frac{\partial^2}{\partial\alpha_i\partial\alpha_j}\langle\buc;\bvc|O|\bub;\bvb\rangle\right|_{\alpha=0}
=\left(\frac{\ell_1(\buc)\ell_3(\bvb)}{\ell_1(\bub)\ell_3(\bvc)}-1\right)\frac{\partial^2}{\partial\alpha_i\partial\alpha_j}&\Theta_{a,b}\left(\buc;\bub|\bvc;\bvb\right)\\
+2\sum_{a_1, b_1}\left(\frac{\ell_1(\buc_{\st})\ell_3(\bvb_{\so})}{\ell_1(\bub_{\st})\ell_3(\bvc_{\so})}-1\right)g(\bvc_{\st},\bvc_{\so})g(\bvb_{\so},\bvb_{\st})&f(\buc_{\so},\buc_{\st})f(\bub_{\st},\bub_{\so})\\
\times f(\bvc_{\so},\buc_{\so})f(\bvb_{\st},\bub_{\st})\frac{\partial}{\partial\alpha_i}\Theta_{a_1, b_2}\left(\bub_{\so};\buc_{\so}|\bvb_{\st};\bvc_{\st}\right)
&\frac{\partial}{\partial\alpha_j}\Theta_{a_2, b_1}\left(\bub_{\st};\buc_{\st}|\bvb_{\so};\bvc_{\so}\right).
\end{split}
\ee

\subsection{Computation of the symmetric form factors\label{FS}}

In order to calculate correlators using \eqref{LeClairMussardoPozsgayHutsaliuk}  the explicit expressions for the symmetric form factors $\mathfrak F_{m,n}$ is required. 

We do not have an efficient algorithm for the calculation of $\mathfrak F_{m,n}$ and $\hat I_{a,b;\:k,l}$, however particular cases
of small particle numbers can be considered. We can apply the formula  \eqref{derive_2}, and the remaining task is to substitute the representation for highest coefficient of correlation function there. These explicit representations of $\Theta^{\alpha}_{m,n}$  were derived in a compact
determinant form in \cite{SLHRP3}:
\be{tONS-ONS}
\begin{split}
\Theta_{m,n}^{\varkappa}(\buc,\bub|\bvb,\bvc)=g(\bub,\bvc)\Delta_g(\bub)\Delta_g(\bvc)\Delta'_g(\buc)\Delta'_g(\bvc)h(\bub,\bub)h(\bvc,\bub)\det \mathcal N.
\end{split}
\ee
Here the block matrix $\mathcal N$ is defined as
\be{N}
\begin{split}
&\mathcal N_{11}=t(\uc_j,\ub_k)\frac{f(\bvb,\ub_k)h(\buc,\ub_k)}{f(\bvc,\ub_k)h(\bub,\ub_k)}+\varkappa_1 t(\ub_k,\uc_j)\frac{h(\ub_k,\buc)}{h(\ub_k,\bub)},\qquad j=1,\dots,m,\qquad k=1,\dots,m,\\
&\mathcal N_{12}=\varkappa_1 t(\vc_k,\uc_j)\frac{h(\vc_k,\buc)}{h(\vc_k,\bub)},\qquad j=1,\dots,m,\qquad k=1,\dots,n,\\
&\mathcal N_{21}=\frac{g(\ub_k,\bvb)}{g(\ub_k,\bvc)}\left(g(\ub_k,\vc_j)+\frac{\varkappa_2/\varkappa_1}{h(\vc_j,\ub_k)}\right), \qquad j=1,\dots,n,\qquad k=1,\dots,m,\\
&\mathcal N_{22}=\delta_{jk}\frac{g(\vc_k,\bvb)}{g(\vc_k,\bvc_k)}\left(1-\varkappa_2\frac{f(\vc_k,\buc)}{f(\vc_k,\bub)}\right),\qquad j=1,\dots,n,\qquad k=1,\dots,n.
\end{split}
\ee
Here we denote $\varkappa_i=e^{\alpha_i}$, where $\alpha$ is defined in \eqref{derive_2}.
Thus, it remains to calculate the sums over partitions in
\eqref{derive_2}. In general this is a complicated problem even for
the algebra symmetry $\mathfrak{gl}(2)$ based models. However, for
small $a, b$ it is easy to obtain the results, since the sums contain only
few terms in these cases.

Further in this section we restrict ourselves to the Gaudin-Yang model. We consider here the case of
repulsion, i.e. coupling constant $c>0$. It was shown by
\cite{TakahashiGas} that there are no string solutions of the Bethe
equations in this case. The case of attractive interaction will be
considered in a separate work since it requires a consideration of the string
solutions.  {Notice, that in comparison with \eqref{betheGY} our Bethe equations \eqref{BEgl3}, \eqref{particles_holes} differ by shift $v\to v+ic/2$. Thus, further in this Section we will assume the shift of $v$ by  $-ic/2$ every time we discuss the Gaudin-Yang model\footnote{Thus, for example, we use $t(v,u)=-\frac{c^2}{(v-u)^2+c^/4}$, $f(v,u)=\frac{v-u+ic/2}{v-u-ic/2}$, etc.}.  We remind that in the Gaudin-Yang model $r_3(v)=e^{iLv}$, $r_1(u)=1$ and $\ell_3(v)=e^{ixv}$, $\ell_1(u)=1$. }

\subsection{Symmetric form factor\label{corr_procedure}}

Consider the symmetric form factors for the correlators of densities. We take the diagonal limit $\vc\to\vb$, $\uc\to \ub$ in \eqref{derive_2} as defined in \eqref{symm_limit}. 

Obviously, symmetric form factor can be expanded into a  Fourier-like series
\be{FA}
\mathfrak F_{a,b}=\sum\left(\frac{\ell_1(\bu_2)\ell_3(\bv_1)}{\ell_1(\bu_1)\ell_3(\bv_2)}-1\right){\mathcal A}^{n,m}_{a,b}\left(\bu_1;\bu_2;\bu_3|\bv_1;\bv_2;\bv_3\right).
\ee
{The sum is taken over partitions $\bu\to\{\bu_1,\bu_2,\bu_3\}$, $\bv\to\{\bv_1,\bv_2,\bv_3\}$ with restrictions $\#\bv_1=\#\bv_2=m$, $\#\bu_1=\#\bu_2=n$.}
We want to stress, that there are terms in \eqref{FA} where $\bv_1=\bv_2$, and thus these terms do not contain oscillating exponents at all.

In models where $\ell_1=1$, such as the Gaudin-Yang model, we are not able to distinguish different ``modes'' in Fourier series. Thus we make the expansion w.r.t. $\ell_3$. Then ${\mathcal  A^n_{a,b}}$ has only one superscript $n=\#\bv$, and the expansion is defined as 
\be{FA1}
\mathfrak F_{a,b}=\sum\left(\frac{\ell_3(\bv^-)}{\ell_3(\bv^+)}-1\right){\mathcal A}^{n}_{a,b}\left(\bu|\bv^+;\bv^-;\bv^0\right).
\ee

\subsection{Symmetric form factor $\mathfrak F_{n,m}$ for $\langle q(x)q(0)\rangle$\label{corr_total}}

We denote Fourier coefficient in front of oscillating parts of correlation function as $I_{a,b}$. One and two particles oscillating parts are given by
\be{I_a01}
I_{0,0}=0,\qquad I_{1,0}=0,\qquad I_{0,1}=0,
\ee
\be{I_a2}
I_{0,2}(\bv)=\mathcal A^1_{0,2}\left(v_1, v_2\right)\left(\ell_3(v_1)/\ell_3(v_2)-1\right)+\mathcal A^1_{0,2}\left(v_2, v_1\right)\left(\ell_3(v_2)/\ell_3(v_1)-1\right),
\ee
where
\be{A02}
\mathcal A^1_{0,2}(v_1,v_2)={ -}2g^2(v_1,v_2),
\ee
and
\be{I_a30}
I_{1,1}(u;v)=0,\qquad I_{2,0}=0.
\ee
Here $\mathcal A^{i}_{k,l}$ are the Fourier coefficients introduced
in \eqref{FA1}.
Three particle oscillating parts are given by
\be{I_a3}
I_{3,0}(\bu)=0,\qquad
I_{0,3}(\bv)=0,\qquad
I_{2,1}(\bu;\bv)=0,
\ee
\be{I_a1b2}
\begin{split}
I_{1,2}\left(\bu;\bv\right)=\mathcal A^1_{1,2}\left(v_1, v_2,u_1\right)\left(\ell_3(v_1)/\ell_3(v_2)-1\right)+\mathcal A^1_{1,2}\left(v_2, v_1,u_1\right)\left(\ell_3(v_2)/\ell_3(v_1)-1\right),
\end{split}
\ee
where
\be{A12}
\mathcal A^1_{1,2}\left(v_1,v_2,u_1\right)=4g^2(v_1,v_2)\frac{g(v_2,u_1)}{h(v_1,u_1)}.
\ee
$I_{3,1}$, $I_{2,2}$, etc. can be calculated too, but the explicit formulas for them become quite lengthy.

One particle $\mathfrak F$ are
\be{Fs_1001}
\mathfrak F_{1,0}=0,\qquad 
\mathfrak F_{0,1}=0.
\ee
Two and three particle form factors are
\be{Fs_02}
\begin{split}
&\mathfrak F_{0,2}=\frac{\partial^2}{\partial x^2}I_{0,2}\\
&+\left.\left.\frac{\partial^2}{\partial x^2}\left\{\sum_{\bv\rightarrow\{v_1,v_2\}} \right.\left(\frac{\ell_3(\vc_1)}{\ell_3(\vb_1)}-1\right)\frac{\partial}{\partial\alpha_2}\Theta_{0,1}\left(\varnothing;\varnothing|\vc_1;\vb_1\right)\frac{\partial}{\partial\alpha_2}\Theta_{0,1}\left(\varnothing;\varnothing|\vc_2;\vb_2\right)\right\}\right|_{\bvc\rightarrow\bvb},
\end{split}
\ee
where the second { (non-oscillating)} term is
\be{Fs_20legend}
\begin{split}
&\left.\frac{\partial^2}{\partial x^2}\left(\frac{\ell_3(\bvc_1)}{\ell_3(\bvb_1)}-1\right)\frac{\partial}{\partial\alpha_2}\Theta_{0,1}\left(\varnothing;\varnothing|\vc_1;\vb_1\right)\frac{\partial}{\partial\alpha_2}\Theta_{0,1}\left(\varnothing;\varnothing|\vc_2;\vb_2\right)\right|_{\bvc\rightarrow\bvb}=4c^2.
\end{split}
\ee
Furthermore
\be{Fs_11}
\begin{split}
\mathfrak F_{1,1}=\frac{\partial^2}{\partial x^2}I_{1,1}=0,\qquad
\mathfrak F_{2,1}=\frac{\partial^2}{\partial x^2}I_{2,1}=0.
\end{split}
\ee
\be{Fs_03}
\begin{split}
\mathfrak F_{0,3}=\frac{\partial^2}{\partial x^2}\left(\frac{\ell_3(\bvc)}{\ell_3(\bvb)}-1\right)\frac{\partial^2}{\partial\alpha^2_2}\Theta_{1,2}\left(\uc;\ub|\bvc;\bvb\right)+\frac{\partial^2}{\partial x^2}I_{0,3}=0,
\end{split}
\ee
and
\be{Fs_12}
\begin{split}
\mathfrak F_{1,2}
=&\frac{\partial^2}{\partial x^2}I_{1,2}\\
+&\frac{\partial^2}{\partial x^2}\sum_{\bv\rightarrow\{\bv_1,\bv_2\}}\left.\left(\frac{\ell_3(\bvc_1)}{\ell_3(\bvb_1)}-1\right)\frac{\partial}{\partial\alpha_2}\Theta_{1,1}(\uc;\ub|\bvc_1;\bvb_1)\frac{\partial}{\partial\alpha_2}\Theta_{0,1}\left(\varnothing;\varnothing|\bvc_2;\bvb_2\right)\right|_{\substack{\buc\rightarrow\bub,\\ \bvc\rightarrow\bvb}},
\end{split}
\ee
where the second { (non-oscillating)} term is
\be{Fs_12legend}
\begin{split}
-4c^2\left\{t(v_1,u_1)+t(v_2,u_1)\right\}.
\end{split}
\ee

We can now argue that the series \eqref{LeClairMussardoPozsgay} converges. We follow again the arguments of \cite{Izergin1984,Korepin1}. We observe, that at large $c$  $\mathfrak F_{a,b}\omega^{(1)}\omega^{(2)}$  are suppressed by the power $c^{2-a-b}$. Using explicitly expressions for $\mathfrak F_{m,n}$ it is easy to estimate now that $n,m$ particle contributions behave as
\be{behaviour}
\left(\frac{B}{c}\right)^n   \left(\frac{Q}{c}\right)^m.
\ee
This shows that the series has good convergence properties if both parameters are small.

\subsection{Symmetric form factor $\mathfrak F_{n,m}$ for $\langle q_{\uparrow}(x)q_{\downarrow}(0)\rangle$}

\label{sec:corr_mixed}

Consider the symmetric form factor for correlators of densities of particles with the different projections of spins.  The oscillating parts with up to two particles are
\be{I_a01ud}
I_{0,0}=0,\qquad I_{1,0}=0\qquad I_{0,1}=0,
\ee
\be{I_a2ud}
I_{2,0}(\bu)=0,\qquad I_{0,2}(\bv)=0,\qquad I_{1,1}(u;v)=0.
\ee
Three particle contributions are given by
\be{I_a3ud}
I_{3,0}(\bu)=0,\qquad
I_{0,3}(\bv)=0,\qquad
I_{2,1}(\bu;\bv)=0,
\ee
\be{I_a1b2ud}
\begin{split}
I_{1,2}\left(\bu;\bv\right)=\mathcal A^1_{1,2}\left(v_1, v_2, u_1\right)\left(\ell_3(v_1)/\ell_3(v_2)-1\right)+\mathcal A^1_{1,2}\left(v_2, v_1,u_1\right)\left(\ell_3(u_2)/\ell_3(u_1)-1\right),
\end{split}
\ee
where
\be{A12ud}
\mathcal A^1_{1,2}\left(v_1,v_2,u_1\right)=2g^2(v_1,v_2)\frac{g(v_2,u_1)}{h(v_1,u_1)}.
\ee
Taking into account terms proportional to $\mathcal J\equiv
\ell_1(\buc)/\ell_1(\bub)-1$ we obtain the symmetric form factors.
\be{Fs_1001ud}
\mathfrak F_{1,0}=0,\qquad 
\mathfrak F_{0,1}=0.
\ee
\be{Fs_02ud}
\begin{split}
\mathfrak F_{0,2}=\frac{\partial^2}{\partial x^2}I_{0,2}=0,\qquad
\mathfrak F_{1,1}=\frac{\partial^2}{\partial x^2}I_{1,1}=0,\qquad
\mathfrak F_{2,1}=\frac{\partial^2}{\partial x^2}I_{2,1}=0.
\end{split}
\ee
\be{Fs_12ud}
\begin{split}
&\mathfrak F_{1,2}=\frac{\partial^2}{\partial x^2}I_{1,2}\\
+&\frac{\partial^2}{\partial x^2}\left\{\sum_{\bv\rightarrow\{\bv_1,\bv_2\}}\left.\left(\frac{\ell_3(\bvc_1)}{\ell_3(\bvb_1)}-1\right)\frac{\partial}{\partial\alpha_1}\Theta_{1,1}\left(\uc;\ub|\bvc_1;\bvb_1\right)\frac{\partial}{\partial\alpha_2}\Theta_{0,1}\left(\varnothing;\varnothing|\bvc_2;\bvb_2\right)\right\}\right|_{\substack{\buc\rightarrow\bub,\\ \bvc\rightarrow\bvb}},
\end{split}
\ee
where the second { (non-oscillating)} term in \eqref{Fs_12ud} is
\be{Fs_12legend_ud}
\begin{split}
-2c^2\left\{t(v_1,u_1)+t(v_2,u_1)\right\}.
\end{split}
\ee
\be{Fs_30}
\mathfrak F_{0,3}=0,\qquad \mathfrak F_{3,0}=0.
\ee
We observe, that $\mathfrak F_{a,b}\omega^{(1)}\omega^{(2)}$ at large $c$ is suppressed by the power $c^{2-a-b}$ in the case of the mixed densities correlator too. 

\subsection{The limit of large imbalance}

\label{densities}

Here we show that if we fix $Q/c$, then the limit of small $B/c$ corresponds to large imbalance.

In the repulsive Gaudin-Yang model the ground state consists of only real roots and the 
densities of Bethe parameters satisfy the system of integral equations \eqref{particles_holes} with $p_1'(x)=0$ and $p_2'(x)=1$.
The densities of fermions are given by
\be{densities_equation1}
\begin{split}
n=n_{\uparrow}+n_{\downarrow}&=\int_{-Q}^Q d\lambda\:\rho_{2}(\lambda),\\
n_{\downarrow}&=\int_{-B}^B d\lambda\:\rho_{1}(\lambda).
\end{split}
\ee
The integral equations  \eqref{particles_holes}  can be solved
iteratively by assuming a small $B/c$ parameter.

In the first approximation the integrals over
$\rho_1(\lambda)$ can be approximated simply by $\rho_1(0)$. This gives
\be{integral_solution1}
\begin{split}
\rho_2(v)&=\frac{1}{2\pi}+2B \frac{1}{2\pi} t(v,0)\rho_1(0),\\    
\rho_1(u)&=-2B \frac{1}{2\pi} K(u,0)\rho_1(0)+
\int_{-Q}^{Q} \frac{dw}{2\pi} t(w,u)\rho_2(w).
\end{split}
\ee
Combining the two equations we get:
\be{integral_solution2}
\begin{split}
\rho_1(u)=\frac{B}{\pi}\rho_1(0) \left[-K(u,0)+\int_{-Q}^{Q}\frac{dw}{2\pi}
t(w,u)t(w,0)\right]+
\frac{1}{2\pi} \int_{-Q}^{Q} \frac{dw}{2\pi}t(w,u).
\end{split}
\ee
Explicit evaluation of the last integral gives
\be{integral_solution}
\rho_1(0)= \frac{1}{\pi^2}\arctan(2Q/c)+O(B/c),
\ee
which in turn leads to
\be{integral_solution3}
n_{\downarrow}=\int_{-B}^{B}du\:\rho_1(u)=\frac{2B}{\pi^2}\arctan(2Q/c)+O(B^2/c^2).
\ee
Regarding the total density we have
\be{integral_solution4}
\begin{split}
n=\frac{Q}{\pi}+ 2B\int_{-Q}^{Q}\frac{dv}{2\pi} t(v,0)\rho_1(0)+O(B^2/c^2)
=\frac{Q}{\pi}+\frac{4B}{\pi^3} \arctan^2(2Q/c)+O(B^2/c^2).
\end{split}
\ee
The last two equations determine $n_{\uparrow/\downarrow}$ including up to first order in $B$. 

Note that these equations are exact in $Q/c$. It is straightforward to perform a further expansion in $Q/s$, if
it is needed for some purpose.

Let us also compute the local 2-body correlations in this
expansion. It follows from the form of the Hamiltonian and the
Hellmann-Feynman theorem that
\begin{equation}
  \vev{ q(0)q(0)}=2\vev{q_\up(0)q_\down(0)}=\frac{1}{L}
    \frac{\partial E_0}{\partial c},
  \end{equation}
  where $E_0$ is the energy of the ground state at fixed particle
  densities.  For the energy density the leading terms in the
  expansion in $B/c$ are known, see for example formula (2.123) of
  \cite{Takahashi-Book} which includes even the $O(B^2/c^2)$
  corrections. The linear term can be computed using the expansion of the
  densities above. In our conventions it reads
  \begin{equation}
    \frac{E_0}{L}=\frac{\pi^2 n^3}{3}-n_\down \left\{
\left(\frac{c^2}{2\pi}+2\pi n^2\right) \tan^{-1}(2\pi n/c) -cn
    \right\}+O(n_\down^2).
  \end{equation}
  Taking a derivative with respect to $c$, while keeping $n$ and
  $n_\down$ fixed we obtain
  \begin{equation}
    \vev{ q(0)q(0)}=2\vev{q_\up(0)q_\down(0)}=
    \frac{8\pi^2 n_\down n^3}{3c^2}+O(n_\down^2).
  \end{equation}
Combining this with the expansion \eqref{integral_solution3}-\eqref{integral_solution4} we see that
\begin{equation}
  \vev{q_\up(0)q_\down(0)}=\frac{32 Q^4B}{3\pi c^3}\times (1+O(B/c)+O(Q^2/c^2)).
\end{equation}

\subsection{Correlation functions}

Now using \eqref{LeClairMussardoPozsgay}, \eqref{Gen_func_gl_21}, \eqref{Gen_func_gl_21_2} and $\mathfrak{F}$, calculated in the section \ref{FS},  we are able to calculate the correlation functions of the highly polarized gas in a strong coupling regime explicitly.

For the first two terms we get
\be{corr_nn}
\begin{split}
\langle q(x)q(0)\rangle=\frac{1}{2\cdot 2!}\int_{-Q}^Q\int_{-Q}^Q\frac{d\bv}{(2\pi)^2}\mathfrak F_{0,2}(\varnothing;\bv)+\frac{1}{2\cdot 2!}\int_{-Q}^Q\int_{-Q}^Q\int_{-B}^B\frac{d\bv du}{(2\pi)^3}\mathfrak F_{1,2}(u;\bv).
\end{split}
\ee
Nontrivial symmetric form factors are
\be{Fs_nn}
\begin{split}
&\mathfrak F_{0,2}(\varnothing;\bv)\omega^{(2)}(\bv)=\left(-2\left(e^{iv_{12}x}+e^{iv_{21}x}\right)+4\right)e^{-8B/(\pi c)}+O(1/c),\\
&\mathfrak F_{1,2}(u;\bv)\omega^{(1)}(\bu)\omega^{(2)}(\bv)=-\frac{16}{c}\left(e^{iv_{12}x}+e^{iv_{21}x}\right)+\frac{32}{c}+O(1/c^2).
\end{split}
\ee
After substitution to  \eqref{corr_nn} we get immediately
\be{correlator_density}
\begin{split}
\langle q(x)q(0)\rangle=\left(\frac{Q^2}{\pi^2}-\frac{\sin^2(Qx)}{x^2\pi^2}\right)e^{-8B/(\pi c)}+\frac{8Q^2B}{\pi^3c}-\frac{8B\sin^2(Qx)}{\pi^3x^2c}+O(c^{-2})\\
=\left(\frac{Q^2}{\pi^2}-\frac{\sin^2(Qx)}{x^2\pi^2}\right)+O(c^{-2}).
\end{split}
\ee
In the same way substituting non trivial symmetric form factors of $\langle q_{\uparrow}(x)q_{\downarrow}(0)\rangle$
\be{Fs_n+n-}
\mathfrak F_{1,2}(u;\bv)\omega^{(1)}(\bu)\omega^{(2)}(\bv)=-\frac{8}{c}\left(e^{iv_{12}x}+e^{iv_{21}x}\right)+\frac{16}{c}+O(1/c^2)
\ee
to \eqref{corr_nn} we get
\be{correlator_flip}
\begin{split}
\langle q_{\uparrow}(x)q_{\downarrow}(0)\rangle=\frac{4Q^2B}{\pi^3c}-\frac{4B\sin^2(Qx)}{\pi^3x^2c}+O(c^{-2}).
\end{split}
\ee

\section*{Conclusion}

In this work we computed the LeClair-Mussardo series for the
correlation functions in models related to algebra symmetries  
$\mathfrak{gl}(2|1)$ and  $\mathfrak{gl}(3)$. We presented two
versions  involving the symmetric and connected form factors. It is important that even though we focused on the particular example of the spin-1/2 Fermions, the expansion theorems are model independent and could be applied in other situations as well.

An important finding is that the symmetric and connected form factors
that enter the LM series do not satisfy the same selection rules as
the on-shell Bethe states  i.e. form factors can be non-zero even when
the corresponding on-shell Bethe states do not exist. In our computations this
is just a result of the algebraic procedure with which we obtained
them. We believe that the physical reason for this is that such form
factors describe multi-particle processes in the presence of the other
particles, therefore it is not required that the same selection rules
should hold as in a situation where the same particles are placed in
the vacuum \cite{Saleur, Takacs1, D22,Pozsgay2}. 

In our concrete example we only treated the ground state of the Gaudin-Yang model, which consists only of real roots. In a future work we also plan to treat cases with string solutions.  This allows to proceed to computations of finite temperature and dynamical correlation functions. Another focus is to study the two-component Bose gas, related to the algebra symmetry 
$\mathfrak{gl}(3)$.

Furthermore, we plan to extend this method to other models. At present there are very few results for correlation functions in spin chains with higher rank symmetries, especially in models with $\mathfrak{o}(N)$ symmetry (see for example \cite{ribeiro2020correlation}). Our method could be helpful in finding new results. Also, it would be interesting to
consider the sine-Gordon model, whose finite volume mean values were investigated in \cite{Takacs-Feher1,Hegedus-SineG-1}. As far as we know, it is not yet clear whether a finite volume expansion theorem exists in that model. Our results could be helpful in settling this issue.

\subsection*{Acknowledgments}

A.H. is grateful to Frank G\"ohmann and Andrii Liashyk for
  numerous discussions. The authors are also thankful to G\'abor
  Tak\'acs and \'Arp\'ad Heged\H{u}s for useful discussions.
A.H. was partially supported by the BME  Nanotechnology and Materials
Science TKP2020 IE grant of NKFIH Hungary  (BME IE-NAT TKP2020). The
authors were also supported by the grant K-16 no. 119204 of NKFIH. B.P.
acknowledges further support from the  the J\'anos Bolyai Research
Scholarship of the Hungarian Academy of Sciences, and from the
\'UNKP-19-4 New National Excellence Program of the Ministry for
Innovation and Technology. Most of this research work was performed
while B.P. was employed by the ``MTA-BME Quantum Dynamics and Correlations Research Group'',
at the Budapest University of Technology and Economics.

\appendix

\section{Appendix\label{quasiReshet}}

Here we give a proof of \eqref{sum_form_final}. We follow the method used in \cite{Izergin1984,Korepin1,ME}.

In models based on the algebra symmetry $\mathfrak{gl}(2|1)$ the off-shell Bethe vectors satisfy the
following co-product property
\be{BV_coprod}
|\bu;\bv\rangle=\sum r_1^{(2)}(\bub_{\so})r^{(1)}_3(\bvb_{\st})f^{-1}(\bvb_{\st},\bub_{\so})\\
 |\bub_{\so};\bvb_{\so}\rangle^{(1)} \otimes|\bub_{\st};\bvb_{\st}\rangle^{(2)}
f(\bub_{\st},\bub_{\so})g(\bvb_{\st},\bvb_{\so}),
\ee
here the superscripts $(i)$, i=1,2 label explicitly that these quantities belong to the
corresponding subsystems. The similar property for a dual Bethe vector is
\be{coprod1}
\langle \bu;\bv|=\sum r_1^{(1)}(\buc_{\st})r^{(2)}_3(\bvc_{\so})f^{-1}(\bvc_{\so},\buc_{\st})\\
 {}^{(1)}\langle\buc_{\so};\bvc_{\so}| \otimes{}^{(2)}\langle\buc_{\st};\bvc_{\st}|
f(\buc_{\so},\buc_{\st})g(\bvc_{\so},\bvc_{\st}).
\ee

The operator $\Qone_i$, $i=1,2$ counts the number of particles of type $i$ in subsystem $1$. Operator $\exp\left(\alpha \Qone\right)$ acting on Bethe vectors of first subsystem produces the factor $\varkappa^{a_1}_1\varkappa_2^{b_1}=\exp(\alpha_1 a_1+\alpha_2b_1)$, where $a_1$, $b_1$ are numbers of particles in the first subsystem, i.e. the eigenvalue of $\Qone_i$. Let us consider simplest example $i=1$, the general case $\exp\left(\alpha_1\Qone_1+\alpha_2\Qtwo_2\right)$ can be proven in the same way. From the co-product property of Bethe vectors in the algebra symmetry $\mathfrak{gl}(2|1)$ based models we can write the following representation  for the (un-normalized) matrix elements of the exponent $\exp\left(\alpha \Qone_1\right)$ 
\be{coprod}
\begin{split}
\langle \buc;\bvc |\exp\left(\alpha \Qone_1\right) |\bub;\bvb \rangle = \sum&\varkappa_1^{a_1} \frac{r_1^{(1)}(\buc_{\st})r_1^{(2)}(\bub_{\so})r_3^{(2)}(\bvc_{\so})r^{(1)}_3(\bvb_{\st})}{f(\bvb,\bub)f(\bvc,\buc)}f(\bvc_{\st},\buc_{\so})f(\bvb_{\so},\bub_{\st})\\
\times S_1\left(\buc_{\so};\bvc_{\so}|\bub_{\so};\bvb_{\so}\right)&S_2\left(\buc_{\st};\bvc_{\st}|\bub_{\st};\bvb_{\st}\right)
f(\buc_{\so},\buc_{\st})f(\bub_{\st},\bub_{\so})g(\bvc_{\so},\bvc_{\st})g(\bvb_{\st},\bvb_{\so}),
\end{split}
\ee
where $S_{\ell}$, $\ell=1,2$ are scalar products of the partial Bethe vectors \eqref{BV_coprod}. 

Substituting  now the scalar products $S_1$, $S_2$ in form \eqref{scal21} into \eqref{coprod}  we get\footnote{In this Appendix we use notation $\bu_{\alpha}$, where $\alpha=1,\dots,4$ are labels of subsets. These subscripts  should not be understand as a shorthand notation \eqref{complement}.}
\be{sum_form1}
\begin{split}
\langle \buc;\bvc |\exp\left(\alpha \Qone_1\right) &|\bub;\bvb \rangle =\sum \varkappa_1^{a_1}\frac{r_1^{(1)}(\buc_{\st})r^{(2)}_1(\bub_{\so})r_3^{(2)}(\bvc_{\so})r_3^{(1)}(\bvb_{\st})}{f(\bvb,\bub)f(\bvc,\buc)}\\
\times f(\bvc_{\st},\buc_{\so})&f(\bvb_{\so},\bub_{\st})f(\buc_{\so},\buc_{\st})f(\bub_{\st},\bub_{\so})g(\bvc_{\so},\bvc_{\st})g(\bvb_{\st},\bvb_{\so})\\
\times &r_1^{(1)}(\bub_1)r_1^{(1)}(\buc_2)r_3^{(1)}(\bvb_1)r^{(1)}_3(\bvc_2)f(\buc_1,\buc_2)f(\bub_2,\bub_1)g(\bvc_2,\bvc_1)g(\bvb_1,\bvb_2)\\
\times &f(\bvc_1,\buc_1)f(\bvb_2,\bub_2){\mathcal Z_{a_1-k_1,n_1}(\buc_2;\bub_2|\bvc_1;\bvb_1)}{\mathcal Z_{k_1,b_1-n_1}(\bub_1;\buc_1|\bvb_2,\bvc_2)}\\
\times &r_1^{(2)}(\bub_3)r_1^{(2)}(\buc_4)r_3^{(2)}(\bvb_3)r_3^{(2)}(\bvc_4)f(\buc_3,\buc_4)f(\bub_4,\bub_3)g(\bvc_4,\bvc_3)g(\bvb_3,\bvb_4)\\
\times &f(\bvc_3,\buc_3)f(\bvb_4,\bub_4){\mathcal Z_{a_2-k_2,n_2}(\buc_4;\bub_4|\bvc_3;\bvb_3)}{\mathcal Z_{k_2,b_2-n_2}(\bub_3;\buc_3|\bvb_4;\bvc_4)},
\end{split}
\ee
where the sets are divided as $\bub_{\so}=\{\bub_1,\bub_2\}$, $\bub_{\st}=\{\bub_3,\bub_4\}$, $\bvb_{\so}=\{\bvb_1,\bvb_2\}$, $\bvb_{\st}=\{\bvb_3,\bvb_4\}$ and in the same way for $\{\buc_{\so}, \buc_{\st}, \bvc_{\so}, \bvc_{\st}\}$.

Further computations are quite bulky but very simple. We want to regroup the factors under the sum
over partition in order to extract the highest coefficients explicitly and express \eqref{sum_form1}
via them. We make some simplification of the long pre-factors separately:
\begin{multline}\label{uc}
f(\buc_1,\buc_3)f(\buc_2,\buc_3)f(\buc_1,\buc_4)f(\buc_2,\buc_4)f(\buc_1,\buc_2)f(\buc_3,\buc_4)\\
=f(\buc_2,\buc_3)f(\buc_1,\buc_4)f(\buc_1,\buc_{23})f(\buc_{23},\buc_4).
\end{multline}
Here $\bu_{\alpha,\beta}=\{\bu_{\alpha},\bu_{\beta}\}$, $\alpha,\beta=1,4$. In the same way we transform factors depending on $\bub_{\alpha}, \bub_{\beta}$. Further, we express all $r_3^{(1)}(\bvc_i)$ via $r_3^{(2)}(\bvc_i)$ and total $r_3(\bvc_i)$, and substitute Bethe equations for $r_3(\bvc_i)$. Thus
\begin{multline}\label{vb}
g(\bvc_1,\bvc_3)g(\bvc_2,\bvc_3)g(\bvc_1,\bvc_4)g(\bvc_2,\bvc_4)g(\bvc_2,\bvc_1)g(\bvc_4,\bvc_3)r_3^{(2)}(\bvc_1)r_3^{(2)}(\bvc_2)r_3^{(1)}(\bvc_2)r_3^{(2)}(\bvc_4)\\
=(-1)^{\#\bv_2\cdot\#\bv_{14}}g(\bvc_{14},\bvc_{23})g(\bvc_2,\bvc_3)g(\bvc_1,\bvc_4)f(\bvc_2,\buc)r_3^{(2)}(\bvc_{14}),
\end{multline}
and we express all $r_3^{(2)}(\bvb_i)$ via $r_3^{(1)}(\bvb_i)$ and total $r_3(\bvb_i)$, and substitute Bethe equations for $r_3(\bvb_i)$
\begin{multline}\label{vc}
g(\bvb_3,\bvb_1)g(\bvb_4,\bvb_1)g(\bvb_3,\bvb_2)g(\bvb_4,\bvb_2)g(\bvb_1,\bvb_2)g(\bvb_3,\bvb_4)r_3^{(1)}(\bvb_3)r_3^{(1)}(\bvb_4)r_3^{(1)}(\bvb_1)r_3^{(2)}(\bvb_3)\\
=(-1)^{\#\bv_3\cdot\#\bv_{14}}g(\bvb_4,\bvb_1)g(\bvb_3,\bvb_2)g(\bvb_{14},\bvb_{23})f(\bvb_3,\bub)r_3^{(1)}(\bvb_{14}).
\end{multline}
Further we collect the rest of the factors
\begin{multline}\label{vcuc}
f(\bvc_3,\buc_1)f(\bvc_4,\buc_1)f(\bvc_3,\buc_2)f(\bvc_4,\buc_2)f(\bvc_1,\buc_1)f(\bvc_3,\buc_3)\\
=f(\bvc_3,\buc_{23})f(\bvc_{14},\buc_1)f(\bvc_3,\buc_1)f(\bvc_4,\buc_2).
\end{multline}
In a similar way simplifications for sets $\{\bub,\bvb\}$ can be performed. Finally we make some simplification of pre-factor with $r_1^{(j)}$, $r_3^{(k)}$:
\begin{multline}\label{r-factor}
r_1^{(1)}(\buc_3)r_1^{(1)}(\buc_4)r_1^{(2)}(\bub_1)r_1^{(2)}(\bub_2)r_1^{(1)}(\bub_1)r_1^{(1)}(\buc_2)r_1^{(2)}(\bub_3)r_1^{(2)}(\buc_4)\\
=r_1(\bub_1)r_1(\buc_4)r_1^{(1)}(\buc_{23})r_1^{(2)}(\bub_{23}).
\end{multline}
Substituting \eqref{uc}--\eqref{r-factor} in the  \eqref{sum_form1} we arrive at
\be{sum_form2}
\begin{split}
\langle \buc;\bvc |\exp\left(\alpha \Qone_1\right) |\bub;\bvb \rangle=\sum \varkappa^{a_1}&\frac{r_1^{(1)}(\buc_{23})r_1^{(2)}(\bub_{23})r_1(\bub_1)r_1(\buc_4)r_3^{(2)}(\bvc_{14})r_3^{(1)}(\bvb_{14})}{f(\bvb,\bub)f(\bvc,\buc)}\\
\times g(\bvb_4,\bvb_1)g(\bvb_2,\bvb_3)g(\bvb_{14},\bvb_{23})&f(\bvb_3,\bub)g(\bvc_{23},\bvc_{14})g(\bvc_3,\bvc_2)g(\bvc_1,\bvc_4)f(\bvc_2,\buc)\\
\times f(\bub_4,\bub_1)f(\bub_3,\bub_2)f(\bub_{23},\bub_1)&f(\bub_4,\bub_{23})f(\buc_2,\buc_3)f(\buc_1,\buc_4)f(\buc_1,\buc_{23})f(\buc_{23},\buc_4)\\
\times f(\bvc_3,\buc_{23})f(\bvc_{14},\buc_1)f(\bvc_3,\buc_1)&f(\bvc_4,\buc_2)f(\bvb_2,\bub_{23})f(\bvb_{14},\bub_4)f(\bvb_1,\bub_3)f(\bvb_2,\bub_4)\\
\times &\mathcal Z_{a_1-k_1,n_1}\left(\buc_2;\bub_2|\bvc_1;\bvb_1\right)\mathcal Z_{k_2,b_2-n_2}\left(\bub_3;\buc_3|\bvb_4,\bvc_4\right)\\
\times&\mathcal Z_{k_1,b_1-n_1}\left(\bub_1;\buc_1|\bvb_2;\bvc_2\right)\mathcal Z_{a_2-k_2,n_2}\left(\buc_4;\bub_4|\bvc_3;\bvb_3\right).
\end{split}
\ee
We collect now the factors that depend on $\{\bvb_1, \bvb_4\}$, $\{\bvc_1, \bvc_4\}$, $\{\bub_2,\bub_3\}$, $\{\buc_2,\buc_3\}$
\begin{multline}\label{HC1}
\sum \varkappa_1^{a_1-k_1}\mathcal Z_{a_1-k_1,n_1}\left(\buc_2;\bub_2|\bvc_1;\bvb_1\right)\mathcal Z_{k_2,b_2-n_2}\left(\bub_3;\buc_3|\bvb_4;\bvc_4\right)f(\bvb_1,\bub_3)f(\bvc_4,\buc_2)\\
\times f(\buc_2,\buc_3)f(\bub_3,\bub_2)g(\bvc_1,\bvc_4)g(\bvb_4,\bvb_1),
\end{multline}
and also write separately factors that depend on $\{\bvb_2, \bvb_3\}$, $\{\bvc_2, \bvc_3\}$, $\{\bub_1,\bub_4\}$, $\{\buc_1,\buc_4\}$
\be{second_HC}
\begin{split}
\sum \varkappa_1^{k_1} r_1(\bub_1)r_1(\buc_4)&\mathcal Z_{k_1,b_1-n_1}\left(\bub_1;\buc_1|\bvb_2;\bvc_2\right)\mathcal Z_{a_2-k_2,n_2}\left(\buc_4;\bub_4|\bvc_3;\bvb_3\right)\\
\times f(\bvb_2,\bub_4)&f(\bvb_{14},\bub_4)f(\bvb_2,\bub_{23}) f(\bub_4,\bub_{123})f(\bub_{23},\bub_1)f(\bvb_3,\bub)g(\bvb_2,\bvb_3)\\
\times &f(\bvc_3,\buc _1)f(\bvc_{14},\buc_1)f(\bvc_3,\buc_{23})f(\buc_{23},\buc_4)f(\buc_1,\buc_{234})f(\bvc_2,\buc)g(\bvc_3,\bvc_2).
\end{split}
\ee
The last expression requires certain simplification. Consider separately the part of factors that depends on $\{\bvb_j,\bub_k\}$
\be{factor_1}
r_1(\bub_1)f(\bvb_2,\bub_{234})f(\bvb_{14},\bub_4)f(\bvb_3,\bub)
=f(\bvb_{23},\bub_{23})f(\bvb,\bub_{14})f(\bvb_3,\bub_1)\frac{f(\bub_1,\bub_{234})}{f(\bub_{234},\bub_1)},
\ee
where we used Bethe equations for $r_1(\bub_1)$ and the part of factors depending on $\{\bvc_j,\buc_k\}$
\be{factor_2}
r_1(\buc_4)f(\bvc_{14},\buc_1)f(\bvc_3,\buc_{123})f(\bvc_2,\buc)
=f(\bvc,\buc_{14})f(\bvc_{23},\buc_{23})f(\bvc_2,\buc_4)\frac{f(\buc_4,\buc_{123})}{f(\buc_{123},\buc_4)}.
\ee
where we used Bethe equations for $r_1(\buc_4)$. {Within \eqref{factor_1}, \eqref{factor_2} the pre-factor of  \eqref{second_HC} can be written as}
\begin{multline}\label{factor}
f(\bvb,\bub_{14})f(\bvc,\buc_{14})f(\bvc_{23},\buc_{23})f(\bvb_{23},\bub_{23})f(\buc_{14},\buc_{23})f(\bub_{14},\bub_{23})\\
\times f(\bvc_2,\buc_4)f(\bvb_3,\bub_1)g(\bvc_3,\bvc_2)g(\bvb_2,\bvb_3)f(\buc_4,\buc_1)f(\bub_1,\bub_4).
\end{multline}
Finally, collecting \eqref{HC1} and simplified \eqref{second_HC} we can conclude our simplification and rewrite \eqref{sum_form2} as
\begin{multline}\label{sum_form_fin}
\langle \buc;\bvc |\exp\left(\alpha \Qone_1\right) |\bub;\bvb \rangle=\sum \frac{r_1^{(1)}(\buc_{23})r_1^{(2)}(\bub_{23})r_3^{(2)}(\bvc_{14})r_3^{(1)}(\bvb_{14})}{f(\bvb,\bub)f(\bvc,\buc)}\\
\times g(\bvb_{14},\bvb_{23})g(\bvc_{23},\bvc_{14})f(\bvb,\bub_{14})f(\bvc,\buc_{14})f(\bvc_{23},\buc_{23})f(\bvb_{23},\bub_{23})f(\buc_{14},\buc_{23})f(\bub_{14},\bub_{23})\\
\times \Theta^{\alpha}_{a_2-k, b_1+n}\left(\bub_{14};\buc_{14}|\bvb_{23};\bvc_{23}\right)
\Theta^{\alpha}_{a_1+k, b_2-n}\left(\bub_{23};\buc_{23}|\bvb_{14};\bvc_{14}\right).
\end{multline}
where $k=k_2-k_1$, $n=n_2-n_1$, and new highest coefficients (HC) $\Theta_{m,n}^{\alpha}$ are defined as
\be{Theta2}
\begin{split}
\Theta^{\alpha}_{m,n}\left(\buc;\bub|\bvc;\bvb\right)=\sum \varkappa_1^k f(\buc_{\st},\buc_{\so})f(\bub_{\so},\bub_{\st})f(\bvb_{\so},\bub_{\so})f(\bvc_{\st},\buc_{\st})&g(\bvc_{\so},\bvc_{\st})g(\bvb_{\st},\bvb_{\so})\\
\times\mathcal Z_{a-k,n}\left(\buc_{\st};\bub_{\st}|\bvc_{\so};\bub_{\so}\right)&\mathcal Z_{k,b-n}\left(\bub_{\so};\buc_{\so}|\bvb_{\st};\bvc_{\st}\right).
\end{split}
\ee
From the comparison with \eqref{scal21} in case when $r_1(\ub)$, $r_1(\uc)$, $r_3(\vb)$ and $r_3(\vc)$ are expressed via the r.h.s. of the Bethe equations system \eqref{BEgl3} it is clear that $\Theta_{m,n}^{\alpha}$ coincides with the scalar product of the on-shell and twisted-on-shell Bethe vectors similarly to the algebra symmetry $\mathfrak{gl}(2)$ related models. Thus we reproduce \eqref{sum_form_final} up to the renumbering of the sets $\bu_{14}\to\bu_{\so}$, $\bu_{23}\to\bu_{\st}$, $\bv_{14}\to\bv_{\so}$, $\bv_{23}\to\bv_{\st}$ and substitution of $r_i^{(2)}=r_i/r_i^{(1)}$, where $r_i$ are expressed via BAE.

\section{Appendix\label{Charge_current}}

In this Appendix we present a simple check of the expansion theorem \eqref{LeClairMussardo}:
we apply it to the conserved charges $I_{\alpha}$ of the models. Here $\alpha$ is an index of the
charge, which typically follows from the expansion of the transfer matrix around special points.
 
In the continuum models the charges are given by the integrals
\be{Charges_ego_sum_conserved}
H_{\alpha}=\int dx \ q_\alpha(x),
\ee
where $q_\alpha(x)$ are the charge density operators\footnote{It is should be noted that in some models, for instance the Lieb-Liniger gas, the real space representations are ill-defined after the third charge \cite{korepin-LL-higher}, and
we expect that similar problems occur also for models of the multi-component gases. Therefore we restrict
ourselves to the well-defined charges, which are the particle densities, the energy density and $H_3$.}. 

It is known that the mean values of the canonical charge density operators are of the form
\be{charge}
\bra{\bar u,\bar v}q_{\alpha}\ket{\bar u,\bar v}=\frac{1}{L}\sum_{i=1}^b h_{\alpha}(v_i),
\ee
where $h_\alpha(v)$ are the one-particle charge eigenvalues. Note that here only the second level
rapidities $\bar v$ contribute; they are also called the momentum-carrying particles. The formula
\eqref{charge} also holds for the total particle density operator $q=q_\uparrow+q_\downarrow$, in
which case $h(v)=1$. In the case of the density $q_\downarrow$ we have formally
\be{density_down_FF}
\bra{\bar u,\bar v}q_\downarrow\ket{\bar u,\bar v}=\frac{1}{L}\sum_{i=1}^a h(u_i),
\quad \text{with} \quad h(u)=1.
\ee
Comparing \eqref{charge} to \eqref{LeClairMussardo} we can compute the symmetric form factors
recursively. These results can then be compared to direct computations of the form factors,
eventually confirming  \eqref{LeClairMussardo}. We consider the simplest cases of this procedure.

\subsection{Symmetric series\label{S_sym}}

Let us denote
\be{rho_def}
\rho(\bu;\bv)=\det G\left(\bu,\bv\right).
\ee
Consider the algebra symmetry $\mathfrak{gl}(2|1)$ related model with
$\partial_v\log r_3(v)=ip'L$, and $r_1=1$. Let us compute explicitly $\mathfrak F$ of
conserved charges for the small values $a$, $b$.

For case $b=1$, $a=0$
\be{FF10}
\mathfrak F_{0,1}(\varnothing,v_1)\rho(\varnothing;\varnothing)+\mathfrak F_{0,0}(\varnothing;\varnothing)\rho(\varnothing;v)=\mathfrak F_{0,1}(\varnothing;v_1)=
\rho(\varnothing;v_1)h_{\alpha}(v_1)/L,
\ee
thus
\be{FS01}
\mathfrak F_{0,1}(\varnothing;v_1)=p'(v_1)h_{\alpha}(v_1).
\ee
For case $b=1$, $a=1$
\be{FF11}
\begin{split}
\mathfrak F_{1,1}(u_1;v_1)\rho(\varnothing;\varnothing)+\mathfrak F_{0,1}(\varnothing;v_1)\rho(u_1;\varnothing)+
&\mathfrak F_{1,0}(u_1;\varnothing)\rho(v_1;\varnothing)+\mathfrak F_{0,0}(\varnothing;\varnothing)\rho(v_1;u_1)\\
&=\rho(v_1;u_1)h_{\alpha}(v_1)/L,
\end{split}
\ee
thus
\be{FS11}
\mathfrak F_{1,1}(u_1;v_1)=t(v_1;u_1)p'(v_1)h_{\alpha}(v_1).
\ee
For case $b=2$, $a=0$  we have similarly
\be{FF02}
\mathfrak F_{0,2}(\varnothing,\bv)\rho(\varnothing;\varnothing)+\mathfrak F_{0,1}\rho(\varnothing;v_1)\rho(\varnothing;v_2)
=\rho(\varnothing;\bv)\left[h_{\alpha}(v_1)+h_{\alpha}(v_2)\right]/L,
\ee
\be{FS02}
\mathfrak F_{0,2}(\varnothing;\bv)=0.
\ee
And for case $b=2$, $a=1$
\be{FF21}
\begin{split}
\mathfrak F_{1,2}(u_1,\bv)&+\mathfrak F_{1,1}(u_1;v_1)\rho(\varnothing;v_2)+\mathfrak F_{1,1}(u_1;v_2)\rho(\varnothing;v_1)
\\+&\mathfrak F_{1,1}(\varnothing;v_2)\rho(u_1;v_1)+\mathfrak F_{1,1}(\varnothing;v_1)\rho(u_1;v_2)=\rho(u_1;\bv)\left[h_{\alpha}(v_1)+h_{\alpha}(v_2)\right]/L.
\end{split}
\ee
Substituting in the last expression
\be{rho_21}
\rho(u_1,\bv)=p'(v_1)p'(v_2)L^2\left(t(v_1,u_1)+t(v_2,u_1)\right)+t(\bv,u_1)L\left(p'(v_1)+p'(v_2)\right),
\ee
we obtain
\be{FS12}
\mathfrak F_{1,2}(u_1;\bv)=t(\bv,u_1)\left(p'(v_1)+p'(v_2)\right)\left(h_{\alpha}(v_1)+h_{\alpha}(v_2)\right).
\ee

Continuing this line of argument every symmetric form factor can be expressed using the charge
eigenvalues, the kernel $t(u,v)$ and some graph theoretical
tools. This leads to a generalization of
the graph theoretical results of  \cite{Pozsgay1} to nested Bethe Ansatz systems. However, further
details of this computation are not important for our present purposes.

Nevertheless we remark that it is also possible to find the symmetric form factors of the current
operators, which describe the 
flow of the charges. This too can be done using the methods of \cite{Pozsgay1}. The resulting form
factors could be used to give an alternative proof of the current mean values in nested systems,
first found in \cite{Pozsgay4}. However, this is beyond the scope of the present paper, and we do
not pursue it.

Instead, let us construct a check of the above formulas, in the the case of the density
operators $q_{\downarrow}(x)$. Taking the first derivative of \eqref{sum_form_final} w.r.t. $\alpha_1$, and
taking into account that $\Theta^{\alpha=0}_{a,b}=\delta_{0,a}\delta_{0,b}$ we obtain the
off-diagonal form factors
\be{rho_gen}
\begin{split}
\langle\buc;\bvc|q_{\downarrow}(x)|\bub,;\bvb\rangle=
  \frac{\partial}{\partial x}&\left.\frac{\partial}{\partial\alpha_1}
  \bra{\buc,\bvc}\exp\left(\alpha \Qone\right) \ket{\bub,\bvb} \right|_{\alpha=0}=\\
&=\frac{\partial}{\partial x}\left(\frac{\ell_1(\buc)\ell_3(\bvb)}{\ell_1(\bub)\ell_3(\bvc)}-1\right)\frac{\partial}{\partial\alpha_1}\Theta^{\alpha=0}_{a, b}\left(\bub;\buc|\bvb;\bvc\right),
\end{split}
\ee
where notation $\exp\left(\alpha \Qone\right)$ is the same as in
\ref{gen_f}, and $\Theta_{a,b}^\alpha$ is given explicitly by \eqref{tONS-ONS}.
Applying the symmetric limit (see \ref{corr_procedure}) we can reproduce \eqref{FS01},
\eqref{FS11}, \eqref{FS12}. This confirms the expansion theorem \eqref{LM} for the specific charge operators.

\subsection{Connected series\label{S_conn}}

Applying \eqref{charge} to \eqref{LeClairMussardo_connected} we can compute connected form factor using the recurrence procedure similar to used in \eqref{S_sym}. Consider again model related to algebra symmetry $\mathfrak{gl}(2|1)$ with $\partial_v\log r_3(v)=ip'L$, and $r_1=1$. 

For case $a=0$, $b=1$ the equation is trivial
\be{FF01c}
\frac{1}{L}h_{\alpha}(v_1)\rho(\varnothing;v_1)=\mathfrak F^c(\varnothing;v_1),
\ee
and we immediately obtain $\mathfrak F_c(\varnothing;v_1)=p'(v_1)h_{\alpha}(v_1)$. In case $b=1$, $a=1$
\be{FF11c}
\frac{1}{L}h_{\alpha}(v_1)\rho(u_1;v_1)=\mathfrak F^c(u_1,v_1)+\mathfrak F^c(\varnothing;v_1)\bar\rho(u_1|v_1).
\ee
It is easy to check that second term in the r.h.s. exactly equal to l.h.s., thus $\mathfrak F^c(u_1,v_1)=0$.  Case $b=1$, $a=2$ is given by
\be{FF21c}
\frac{1}{L}h_{\alpha}(v_1)\rho(\bu;v_1)=\mathfrak F^c(\bu;v_1)+\mathfrak F^c(\varnothing;v_1)\bar\rho(\bu|v_1).
\ee
Again, the second term in the r.h.s. is equal to 
\be{FF21c_rhs}
p'(v_1)h_{\alpha}(v_1)\left\{t(v_1,u_1)t(v_1,u_2)-K(u_1,u_2)\left(t(v_1,u_1)+t(v_1,u_2)\right)\right\}
\ee
and it coincides with the l.h.s. of  \eqref{FF21c}, thus $\mathfrak F^c(\bu;v_1)=0$. Continuing this procedure it is easy to check that $\mathfrak F^c(\bu;v)=0$ for the case $\#\bu\ge1$.

In a case $b=2$, $a=0$ we have
\be{FF02c}
\rho(\varnothing;\bv)\frac{h_{\alpha}(v_1)h_{\alpha}(v_2)}{L}=\mathfrak F^c(\varnothing;\bv)+\mathfrak F^c(\varnothing;v_1)\bar\rho(v_2|v_1)+\mathfrak F^c(\varnothing;v_1)\bar\rho(v_1|v_2),
\ee
taking into account that $\rho(\varnothing;\bv)=L^2h_{\alpha}(v_1)h_{\alpha}(v_2)$ and using results of \eqref{FF01c}--\eqref{FF21c} it is easy to obtain that $\mathfrak F^c(\varnothing;\bv)=0$. 

In case $b=2$, $a=1$ we have
\be{FF12c}
\rho(u_1;\bv)\frac{h_{\alpha}(v_1)+h_{\alpha}(v_2)}{L}=\mathfrak F^c(u_1;\bv)+\mathfrak F^c(\varnothing;v_1)\bar\rho(u_1;v_2|v_1)+\mathfrak F^c(\varnothing;v_2)\bar\rho(u_1;v_1|v_2),
\ee
and using \eqref{rho_21} and $\bar\rho(u_1;v_1|v_2)=p'(v_1)L[t(v_1,u_1)+t(v_2,u_1)]+t(v_1,u_1)t(v_2,u_2)$ we arrive at
\be{Fc12}
\mathfrak F^c(u_1;\bv)=t(v_1,u_1)t(v_2,u_1)\left[p'(v_1)h_{\alpha}(v_2)+p'(v_2)h_{\alpha}(v_1)\right].
\ee
Continuing a similar procedure we can calculate further $\mathfrak F^c$ and even establish a general graph rule, similarly to case of symmetric form factors \cite{Pozsgay1}. Again using \eqref{rho_gen} and applying to this formula corresponding symmetric limit \eqref{connected_limit} we can easily reproduce results that follow from \eqref{FF01c}--\eqref{FF12c}. This confirms the expansion \eqref{LeClairMussardo_connected} for the charge operator.

\section{Appendix\label{LM_vs_IK}}

Correlation function of densities in the Lieb-Liniger model was computed in \cite{Izergin1984,Korepin1}. 
\be{Q_0}
\begin{split}
\langle q(x)q(0)\rangle=\frac{Q^2}{\pi^2}+\frac{4Q^3}{c\pi^3}&-\left(1+\frac{2}{c}\frac{\partial}{\partial x_r}\right)\frac{\sin^2(Qx_r)}{\pi^2 x_r^2}-\frac{8Q}{\pi c}\frac{\sin^2(Qx)}{\pi^2x^2}\\
&+\frac{16}{c}\frac{\partial}{\partial x}\frac{\sin(Qx)}{x}\int_{-Q}^Q\frac{du}{(2\pi)^3}\sin(ux)\ln\left(\frac{Q+u}{Q-u}\right)+O(Q^2/c^2),
\end{split}
\ee
with $x_r=x(1+2Q/\pi c)$.

Here we reproduce this result using LM-theorem. {The Lieb-Liniger model is related to the algebra symmetry $\mathfrak{gl}(2)$.}
Using \eqref{LeClairMussardoPozsgay} and \eqref{Gen_func_gl_21} we  arrive at
\be{LeClairMussardoPozsgay2b}
\langle q(x)q(0)\rangle=\frac{1}{2}\frac{\partial^2}{\partial x^2}\sum_{k=0}^{\infty}\frac{1}{k!}\int_{-Q}^Q\frac{d\bar t}{(2\pi)^k}\;\omega(\bar t) I_k(\bar t),\qquad \omega(u)=\frac{1}{c}\exp\left(-\frac{1}{2\pi c}\int_{-Q}^Q K(u,x)dx\right),
\ee
where $I_k$ are irreducible parts for algebra symmetry $\mathfrak{gl}(2)$ related models and $\omega$ is a weight function.

Oscillating parts of symmetric form factor in the algebra symmetry $\mathfrak{gl}(2)$ related  models are
\be{I2_I3_gl2}
\begin{split}
&I_{2}(\bu)=\mathcal A^1_{2}\left(u_1, u_2\right)\left(\ell_1(u_1)/\ell_1(u_2)-1\right)+\mathcal A^1_{2}\left(u_2, u_1\right)\left(\ell_1(u_2)/\ell_1(u_1)-1\right),\\
&I_{3}(\bu)=\sum_{\mathbb P}\mathcal A^1_{3}\left(u_{p_1}, u_{p_2},u_{p_3}\right)\left(\ell_1(\bu_{p_1})/\ell_1(\bu_{p_2})-1\right)+x\sum_{\mathbb P}\tilde {\mathcal A}^1_3(u_{p_1},u_{p_2})  \ell_1(u_{p_1})/\ell_1(u_{p_2}),\\
\end{split}
\ee
where $\ell_1(u)=iLx$ in the Lieb-Liniger gas and we denote Fourier coefficients of the symmetric form factors as $\mathcal A^k_n$.  The sum is taken over all permutations $\mathbb P$ of spectral parameters $\{u_1, u_2, u_3\}$.
\be{Fourier_limit_gl2}
\begin{split}
&\mathcal A^1_2(u_1,u_2)=2g^2(u_1,u_2)\left(1+\frac{2u_{12}}{ic}\right)+O(c^{1}),\\
&\mathcal A^1_3(u_1,u_2,u_3)=-8g^2(u_1,u_2)\left(\frac{u_{32}}{u_{31}}+\frac{u_{31}}{u_{32}}\right)+O(c^1),\qquad \tilde{\mathcal A}^1_3(u_1,u_2)=4icg(u_1,u_2).
\end{split}
\ee
Total $\mathfrak F$ are
\be{Fourier_limit_gl2c}
\mathfrak F_2(\bu)=\frac{\partial^2}{\partial x^2}I_2+4c^2+O(c^{1}),\qquad
\mathfrak F_3(\bu)=\frac{\partial^2}{\partial x^2}I_3+16\cdot 3!c^2+O(c^1).
\ee

We denote $n$-particle contributions as $\Gamma_n$ and take into account terms up to the order $O(c^{-1})$. In the two-particles contribution we should take into account  factor $\omega(u)=c^{-1}\exp(-4Q/(\pi c))+O(c^{-2})$. We compute the oscillating parts separately 
\be{corr_density_c_1_2p}
\begin{split}
\Gamma_2=2\int_{-Q}^Q\int_{-Q}^Q\frac{du_1du_2}{(2\pi)^2 2\cdot 2!}&
\left[e^{iu_{12}x}\left(1+\frac{2u_{12}}{ic}\right)+e^{iu_{21}x}\left(1+\frac{2u_{21}}{ic}\right)\right]e^{-\frac{4Q}{\pi c}}\\
&=-\left(1+\frac{2}{c}\frac{\partial}{\partial x}\right)\frac{\sin^2(Qx)}{\pi^2 x^2}+\frac{4Q}{\pi c}\frac{\sin^2(Qx)}{\pi^2x^2}+O(c^{-2}).
\end{split}
\ee
In the same way we calculate the three particle contributions that have order $1/c$
\be{corr_density_c_1_3p}
\begin{split}
&\Gamma_3=-\frac{8Q}{c\pi^3}\frac{\sin^2(Qx)}{x^2}+\frac{2}{c}\frac{\partial}{\partial x}\left[\sin(Qx)\int_{-Q}^Q\frac{du}{(2\pi)^3}\sin(ux)\ln\left(\frac{Q+u}{Q-u}\right)\right]+O(c^{-2}),\\
&\tilde\Gamma_3=-\frac{4Q}{\pi^3c}\frac{\sin^2(Qx)}{x^2}
-\frac{2Qx}{c}\frac{\partial}{\partial x}\frac{\sin^2(Qx)}{\pi^3 x^2 c}+O(c^{-2}).
\end{split}
\ee
Collecting \eqref{corr_density_c_1_2p} and \eqref{corr_density_c_1_3p} and expanding $x_r$ in the oscillating part of \eqref{Q_0} up to $1/c$ it is easy to convince that oscillating part is reproduced form LM-series.
Finally, the constant parts coming from the 2- and 3-particle contributions are
\be{non_osc_Lm_g2}
\frac{Q^2}{\pi^2}+\frac{4Q^3}{c\pi^3},
\ee
that exactly coincides with the constant term in \eqref{Q_0}.

\addcontentsline{toc}{section}{References}

\providecommand{\href}[2]{#2}\begingroup\raggedright\endgroup

\end{document}